\documentclass[11pt,preprint,trackchanges]{aastex63}

\usepackage{url}
\def\apjl{\rm ApJL}
\def\apjs{\rm ApJS}

\def\aap{\rm AAP}

\newcommand{\gtorder}{\mathrel{\raise.3ex\hbox{$>$}\mkern-14mu
            \lower0.6ex\hbox{$\sim$}}}
\newcommand{\ltorder}{\mathrel{\raise.3ex\hbox{$<$}\mkern-14mu
            \lower0.6ex\hbox{$\sim$}}}

\shorttitle{The Radius of PSR~J0740$+$6620}
\shortauthors{Miller, Lamb, Dittmann, et al.}

\begin{document}

\title{THE RADIUS OF PSR J0740$+$6620 FROM NICER AND XMM-NEWTON DATA
}

\correspondingauthor{M.~C.~Miller}
\email{miller@astro.umd.edu}

\author[0000-0002-2666-728X]{M.~C.~Miller}
\affiliation{Department of Astronomy and Joint Space-Science Institute, University of Maryland, College Park, MD 20742-2421 USA}

\author[0000-0002-3862-7402]{F.~K.~Lamb}
\affiliation{Illinois Center for Advanced Studies of the Universe and Department of Physics, University of Illinois at Urbana-Champaign, 1110 West Green Street, Urbana, IL 61801-3080, USA}
\affiliation{Department of Astronomy, University of Illinois at Urbana-Champaign, 1002 West Green Street, Urbana, IL 61801-3074, USA}

\author[0000-0001-6157-6722]{A.~J.~Dittmann}
\affiliation{Department of Astronomy and Joint Space-Science Institute, University of Maryland, College Park, MD 20742-2421 USA}

\author[0000-0002-9870-2742]{S.~Bogdanov}
\affiliation{Columbia Astrophysics Laboratory, Columbia University, 550 West 120th Street, New York, NY 10027, USA}

\author{Z.~Arzoumanian}
\affiliation{X-Ray Astrophysics Laboratory, NASA Goddard Space Flight Center, Code 662, Greenbelt, MD 20771, USA}

\author[0000-0001-7115-2819]{K.~C.~Gendreau}
\affiliation{X-Ray Astrophysics Laboratory, NASA Goddard Space Flight Center, Code 662, Greenbelt, MD 20771, USA}

\author[0000-0002-6449-106X]{S.~Guillot}
\affil{IRAP, CNRS, 9 avenue du Colonel Roche, BP 44346, F-31028 Toulouse Cedex 4, France}
\affil{Universit\'{e} de Toulouse, CNES, UPS-OMP, F-31028 Toulouse, France}

\author[0000-0002-6089-6836]{W.~C.~G.~Ho}
\affiliation{Department of Physics and Astronomy, Haverford College, 370 Lancaster Avenue, Haverford, PA 19041, USA}

\author[0000-0002-5907-4552]{J.~M. Lattimer}
\affiliation{Department of Physics and Astronomy, Stony Brook University, Stony Brook, NY 11794-3800, USA}

\author[0000-0002-1661-4029]{M.~Loewenstein}
\affiliation{Department of Astronomy, University of Maryland, College Park, MD 20742-2421 USA}
\affiliation{X-Ray Astrophysics Laboratory, NASA Goddard Space Flight Center, Code 662, Greenbelt, MD 20771, USA}
\affiliation{Center for Research and Exploration in Space Science and Technology, NASA/GSFC, Greenbelt, MD 20771}

\author[0000-0003-4357-0575]{S.~M.~Morsink}
\affiliation{Department of Physics, University of Alberta, 4-183 CCIS, Edmonton, AB T6G 2E1, Canada}

\author[0000-0002-5297-5278]{P.~S.~Ray}
\affiliation{Space Science Division, U.S. Naval Research Laboratory, Washington, DC 20375, USA}

\author[0000-0002-4013-5650]{M.~T.~Wolff}
\affiliation{Space Science Division, U.S. Naval Research Laboratory, Washington, DC 20375, USA}

\author{C.~L.~Baker}
\affiliation{Applied Engineering and Technology Directorate, NASA Goddard Space Flight Center, Greenbelt, MD 20771, USA}

\author{T.~Cazeau}
\affiliation{X-Ray Astrophysics Laboratory, NASA Goddard Space Flight Center, Code 662, Greenbelt, MD 20771, USA}

\author{S.~Manthripragada}
\affiliation{Instrument Systems and Technology Division, NASA Goddard Space Flight Center, Greenbelt, MD 20771, USA}

\author[0000-0001-9803-3879]{C.~B.~Markwardt}
\affiliation{Astrophysics Science Division, NASA Goddard Space Flight Center, Greenbelt, MD 20771, USA}

\author{T.~Okajima}
\affiliation{X-Ray Astrophysics Laboratory, NASA Goddard Space Flight Center, Code 662, Greenbelt, MD 20771, USA}

\author{S.~Pollard}
\affiliation{X-Ray Astrophysics Laboratory, NASA Goddard Space Flight Center, Code 662, Greenbelt, MD 20771, USA}

\author[0000-0002-1775-9692]{I.~Cognard}
\affiliation{Station de Radioastronomie de Nan\c{c}ay, Observatoire de Paris, CNRS/INSU, Universit{\'e} d’Orl{\'e}ans, 18330, Nan\c{c}ay, France}
\affiliation{9 Laboratoire de Physique et Chimie de l’Environnement, CNRS, 3A Avenue de la Recherche Scientifique, 45071, Orl{\'e}ans Cedex 2, France}

\author[0000-0002-6039-692X]{H.~T.~Cromartie}
\affiliation{Cornell Center for Astrophysics and Planetary Science and Department of Astronomy, Cornell University, Ithaca, NY 14853, USA}
\affiliation{NASA Hubble Fellowship Program Einstein Postdoctoral Fellow}

\author[0000-0001-8384-5049]{E.~Fonseca}
\affiliation{Department of Physics, McGill University, 3600 rue University, Montr\'eal, QC H3A 2T8, Canada}
\affiliation{McGill Space Institute, McGill University, 3550 rue University, Montr\'eal, QC H3A 2A7, Canada}
\affiliation{Department of Physics and Astronomy, West Virginia University, P.O. Box 6315, Morgantown, WV 26506, USA}
\affiliation{Center for Gravitational Waves and Cosmology, West Virginia University, Chestnut Ridge Research Building, Morgantown, WV 26505, USA}

\author[0000-0002-9049-8716]{L.~Guillemot}
\affiliation{Station de Radioastronomie de Nan\c{c}ay, Observatoire de Paris, CNRS/INSU, Universit{\'e} d’Orl{\'e}ans, 18330, Nan\c{c}ay, France}
\affiliation{9 Laboratoire de Physique et Chimie de l’Environnement, CNRS, 3A Avenue de la Recherche Scientifique, 45071, Orl{\'e}ans Cedex 2, France}

\author[0000-0002-0893-4073]{M.~Kerr}
\affiliation{Space Science Division, Naval Research Laboratory, Washington, DC 20375 5352, USA}

\author[0000-0002-4140-5616]{A.~Parthasarathy}
\affiliation{Max-Planck-Institut f{\"u}r Radioastronomie, Auf dem H{\"u}gel 69, D-53121 Bonn, Germany}

\author[0000-0001-5465-2889]{T.~T.~Pennucci}
\affiliation{National Radio Astronomy Observatory, 520 Edgemont Road, Charlottesville, VA 22903, USA}
\affiliation{Institute of Physics, E{\"o}tv{\"o}s Lor{\'a}nd University, P{\'a}zm{\'a}ny P.s. 1/A, 1117 Budapest, Hungary}

\author[0000-0001-5799-9714]{S.~Ransom}
\affiliation{National Radio Astronomy Observatory, 520 Edgemont Road, Charlottesville, VA 22903, USA}

\author[0000-0001-9784-8670]{I.~Stairs}
\affiliation{Department of Physics and Astronomy, University of British Columbia, 6224 Agricultural Road, Vancouver, BC V6T 1Z1, Canada}

\begin{abstract}

PSR~J0740$+$6620 has a gravitational mass of $2.08\pm 0.07~M_\odot$, which is the highest reliably determined mass of any neutron star.  As a result, a measurement of its radius will provide unique insight into the properties of neutron star core matter at high densities.  Here we report a radius measurement based on fits of rotating hot spot patterns to \textit{Neutron Star Interior Composition Explorer} (\textit{NICER}) and \textit{X-ray Multi-Mirror} (\textit{XMM-Newton}) X-ray observations.  We find that the equatorial circumferential radius of PSR~J0740$+$6620 is $13.7^{+2.6}_{-1.5}$~km (68\%).  We apply our measurement, combined with the previous \textit{NICER} mass and radius measurement of PSR~J0030$+$0451, the masses of two other $\sim 2~M_\odot$ pulsars, and the tidal deformability constraints from two gravitational wave events, to three different frameworks for equation of state modeling, and find consistent results at $\sim 1.5-3$~times nuclear saturation density.  For a given framework, when all measurements are included the radius of a $1.4~M_\odot$ neutron star is known to $\pm 4$\% (68\% credibility) and the radius of a $2.08~M_\odot$ neutron star is known to $\pm 5$\%. The full radius range that spans the $\pm 1\sigma$ credible intervals of all the radius estimates in the three frameworks is $12.45\pm 0.65$~km for a $1.4~M_\odot$ neutron star and $12.35\pm 0.75$~km for a $2.08~M_\odot$ neutron star.

\end{abstract}

\keywords{dense matter --- equation of state --- neutron star --- X-rays: general}

\section{INTRODUCTION}
\label{sec:introduction}

Neutron stars are unique laboratories for the study of dense matter.  Their cores consist of matter that is believed to be catalyzed to the ground state, at a few times nuclear saturation density (a mass density $\rho_s\approx 2.7-2.8\times 10^{14}~{\rm g~cm}^{-3}$, or a baryonic number density $n_s\approx 0.16$~fm$^{-3}$).  The combination of high density and the expected large neutron-proton asymmetry in neutron star cores cannot be duplicated in laboratories.  Hence, observations of neutron stars can provide us with a valuable window into an otherwise inaccessible realm of nuclear physics.

Over the last several years great strides have been made in neutron star observations, and thus in our understanding of the equation of state (EOS: pressure as a function of energy density) of neutron star matter at high densities (see, e.g.,  \citealt{1997ApJ...490L..91P,2005ApJ...619..483B,2010ApJ...722...33S,2013arXiv1312.0029M,2016EPJA...52...63M,2016ApJ...820...28O,2017A&A...608A..31N} for earlier perspectives).  Three neutron stars have a gravitational mass established to be $M\sim 2~M_\odot$: PSR~1614--2230 with $M=1.908\pm 0.016~M_\odot$ (the uncertainties here and below are for the 68\% credible region) \citep{2010Natur.467.1081D,2016ApJ...832..167F,2018ApJ...859...47A}; PSR~J0348+0432 with $M=2.01\pm 0.04~M_\odot$ \citep{2013Sci...340..448A}; and PSR~J0740$+$6620 with $M=2.08\pm 0.07~M_\odot$ (\citealt{2020NatAs...4...72C,2021arXiv210400880F}).  The existence of such high-mass neutron stars indicates that the EOS of neutron star matter is relatively hard, i.e., that it yields high pressures at a few times $\rho_s$.  The lack of a clear signature of tidal deformation in gravitational wave observations of GW170817 \citep{2017PhRvL.119p1101A,2018PhRvL.121p1101A,2018PhRvL.121i1102D} and GW190425 \citep{2020ApJ...892L...3A} indicates that the EOS is not too hard.  When taken together, these high measured masses and the upper limits on the tidal deformability derived from these gravitational wave observations have already narrowed significantly the range of allowed EOS models.  The neutron star radius and mass measurements made using data from the \textit{Neutron Star Interior Composition Explorer} (\textit{NICER}) are potentially even more informative.  

The first simultaneous measurements of the mass and radius of a neutron star made using \textit{NICER} data were those of the millisecond pulsar PSR~J0030$+$0451, which was determined to have a gravitational mass of $M \approx 1.44 \pm 0.15~M_\odot$ and an equatorial circumferential radius of $R_e \approx 13 \pm 1$~km (\citealt{2019ApJ...887L..24M,2019ApJ...887L..21R}; compare with the $11.9\pm 1.4$~km radius inferred for the two $\sim 1.4~M_\odot$ neutron stars in the gravitational wave event GW170817 [\citealt{2018PhRvL.121p1101A,2018PhRvL.121i1102D}]).  Assuming that systematic errors in the \textit{NICER} measurements are unimportant, and hence that the fractional uncertainty of each of these measurements decreases as the inverse square root of the observing time devoted to each pulsar, we expect these uncertainties to become $\sim 30$\%–-40\% smaller within the next few years.  A precision this high will improve significantly our understanding of neutron star matter. It is also important to determine whether the masses and radii, and the EOS of neutron star matter, determined using \textit{NICER} observations of other neutron stars are consistent with those determined using \textit{NICER} observations of PSR~J0030$+$0451.

In particular, it is valuable to measure the radii of higher-mass neutron stars.  Such stars have higher central densities than a $\sim 1.4~M_\odot$ star, which means that a radius measurement will probe the EOS in a higher-density regime.  For example, whereas the measurement of the radius of a $1.4~M_\odot$ star has its greatest impact on our understanding of matter at 1.6 times nuclear saturation density, the measurement of the radius of a $2.0~M_\odot$ star tells us primarily about matter at 2.2 times nuclear saturation density (see Section IV.D of \citealt{2020arXiv200906441D}, and also see \citealt{2020ApJ...899....4X} for additional perspectives on the importance of radius measurements for high-mass neutron stars).  As pointed out in the context of the $\sim 1.9~M_\odot$ pulsar PSR~J1614$-$2230, even a low-precision radius measurement of a high-mass pulsar, or indeed even a solid lower limit to the radius, will be useful in the construction of more accurate EOS for dense matter \citep{2016ApJ...822...27M}.

Here we report our analysis of X-ray data for the $\sim 2.1~M_\odot$ pulsar PSR~J0740$+$6620, which has a rotational frequency of 346.53~Hz \citep{2020NatAs...4...72C}.  In contrast to PSR~J0030$+$0451, which is an isolated neutron star, PSR~J0740$+$6620 is in a binary and therefore we have independent measurements of the mass and orbital inclination.  This improves the precision of the fits.  The challenge presented by PSR~J0740$+$6620 is that its \textit{NICER} count rate is only $\sim 5$\% that of PSR~J0030$+$0451.  As a result, counts in \textit{NICER} data from PSR~J0740$+$6620 are dominated by sources unassociated with the pulsar, to a far greater degree than is true for PSR~J0030$+$0451.  This significantly reduces the modulation fraction and harmonic structure in the X-ray pulse waveform, which in turn reduces the information we have about the emitting regions (which we henceforth call ``spots") and the mass and radius of the star.  If we had precise and reliable information about the background in \textit{NICER} observations we could use this to improve our mass and radius estimates, but at present we need to turn to other sources of information.

In this paper we therefore present an analysis of PSR~J0740$+$6620 based on both \textit{NICER} and \textit{X-ray Multi-Mirror} (\textit{XMM-Newton}) data.  The \textit{NICER} data provide information about the modulated emission from the star, which is produced from hot spots on the star as the star rotates.  The \textit{XMM-Newton} data have a far smaller rate of background counts than the \textit{NICER} data and are therefore informative about the total flux from the star.  Combined, the \textit{NICER} and \textit{XMM-Newton} data sets thus yield an improved measurement of the modulation fraction and the harmonic structure from PSR~J0740$+$6620, and hence improve the precision with which we can measure the radius.  In particular, stronger gravitational lensing allows more of the star to be visible, which lowers the modulation of the signal. Since gravitational lensing depends on the ratio of mass to radius, for fixed mass, an increase in the pulsed fraction means that a larger radius will be inferred if all other parameters are fixed.

In Section~\ref{sec:observations} we describe the selection of the \textit{NICER} and \textit{XMM-Newton} data, including blank-sky \textit{XMM-Newton} data that have been accumulated over the duration of the \textit{XMM-Newton} mission and that help us refine further our estimate of the stellar flux.  In Section~\ref{sec:methods} we discuss our analysis methods.  These are based on the approaches described in previous \textit{NICER} papers \citep{2019ApJ...887L..26B,2019ApJ...887L..24M,2021arXiv210406928B}, but are augmented here by the extra data sets and our inclusion of background information from the \textit{XMM-Newton} blank-sky observations.  In Section~\ref{sec:fits-to-data} we present our analysis results including tests of the adequacy of our models; we find that the radius of PSR~J0740$+$6620 is $R_e=13.7^{+2.6}_{-1.5}$~km at 68\% credibility.  We also describe several differences between our analysis and the analysis in the parallel paper \citet{2021..............R}, in Section~\ref{sec:differences}.  In Section~\ref{sec:EOS} we present the implications of our PSR~J0740$+$6620 analysis, combined with our previous results on PSR~J0030+0451 and other data, for the EOS of cold dense matter.  Our conclusions are in Section~\ref{sec:conclusions}.   We present our full corner plots and tables of posteriors in the appendices.  Samples from our full posterior probability distributions are available at https://zenodo.org/record/4670689.

\section{OBSERVATIONS}
\label{sec:observations}

\subsection {Data Used}
\label{sec:data-used}

We base our analysis of PSR J0740+6620 on \textit{NICER} event data that included individual event times, event energies, and event pulsation phase. Our observing strategy and data reduction resulted in a data set with very low noise properties, as was necessary because of the faintness of this pulsar. 

Our analysis of the X-ray data on PSR~J0740$+$6620 comes from a series of 215 observations obtained with \textit{NICER}\ XTI collected between 2018 September 21 and 2020 April 17, with a net exposure time of 1602.68\,ksec (after all filtering  was applied, as described below).  The final event data was obtained using the following filtering criteria: (i) we only included ObsIDs when all 52 \textit{NICER} detectors are active; (ii) we excluded events with energies outside the range 0.3 to 3.0 keV; (iii) we excluded events from DetID 34 (which frequently exhibits elevated count rates compared to the average); (iv) we only considered time intervals when PSR~J0740$+$6620 was situated at least $80^{\circ}$ from the Sun to minimize optical loading and scattered solar X-ray background photons; and (v) we rejected data obtained at low cut-off rigidities (COR\_SAX$<5$) to minimize particle background contamination.  The filtered events were assigned pulse phases using the PSR~J0740$+$6620 ephemeris from \citet{2021arXiv210400880F}.  This event list ultimately consisted of 7334 separate good time intervals (GTIs).  We then searched this event list for the pulsar signal.  If the events in a particular GTI did not increase the pulsar signal to noise we rejected that GTI.  The pulsar detection significance was highest in the energy range 0.31 to 1.18 keV.  Ultimately, after all filtering and other exclusions were finished, we ended up with 521004 \textit{NICER}\ XTI events.  The details of the \textit{NICER} observations and pulse detection for PSR~J0740$+$6620 can be found in \citet{2021..............W}.

We use the latest available calibration products, namely the nixtiref20170601v002 redistribution matrix file (RMF) and the nixtiaveonaxis20170601v004 ancillary response file (ARF).  For the latter, the effective areas per energy channel were corrected by a factor of $51/52$ to account for the removal of all events from DetID 34.  The response properties of \textit{NICER} incorporated in these calibration products represented the best effort by the instrument modeling team that were available in August 2020 when the analysis presented in this paper was begun.

PSR~J0740$+$6620 was also targeted with \textit{XMM-Newton} on three separate occasions: 2019 October 26 (ObsID 0851181601), 2019 October 28 (ObsID 0851181401), and 2019 November 1 (ObsID 0851181501) as part of a Director's Discretionary Time program.  All three European Photon Imaging Camera (EPIC) instruments (pn, MOS1, and MOS2) were operated in 'Full Frame' imaging mode with the 'Thin' filters in the optical path.  Due to the long detector read-out times (73.4~ms for EPIC-pn and 2.6~s for EPIC-MOS1/2), the data do not permit a pulse-phase-resolved analysis.  Data summed over phase are used in the analysis that follows.  The \textit{XMM-Newton} data were processed with the Science Analysis Software (SAS\footnote{The \textit{XMM-Newton} SAS is  developed and maintained by the Science Operations Centre at the European Space Astronomy Centre and the Survey Science Centre at the University of Leicester.}) using the recommended procedures.  We employed the recommended event grade PATTERN and FLAG values that ensured only real photon events recorded by the instruments were included in our analysis ($\leq 12$ for MOS1/2 and $\leq 4$ for pn). All three of these \textit{XMM-Newton} observations were obtained while the satellite was in a part of its orbit that suffered from intense particle flaring effects.  Only 6.8 ksec of the total 20.8 ksec of EPIC-pn exposure had sufficiently low background to be used in our analysis, and only 18.0/18.7 ksec of the total 26.7/34.3 ksec of MOS1/2 exposures could be utilized.  A circular extraction region of radius 25" centered on the radio timing position of PSR~J0740$+$6620 was used to produce clean source event lists.  The \textit{XMM-Newton} instrument response files specific to the PSR J0740+6620 observations were then generated using the rmfgen and arfgen tools in SAS.

	Under normal observing circumstances, one uses regions nearby to the target source in the instrument image plane from the same observations to estimate the observational background. In our case, however, due to the short durations of these observations, very few background counts are available to estimate the background. As an alternative, we obtained representative background estimates with longer exposures from the blank sky event files provided by the \textit{XMM-Newton} Science Operations Centre. The blank sky images were filtered in the same manner as the PSR~J0740$+$6620 images and the background was extracted from the same location on the detector image plane as the pulsar. The resulting background spectrum was then rescaled so that the exposure times and pixel-area factors match those of the PSR~J0740$+$6620 exposures.

\subsection {Calibration of the Instruments}
\label{sec:calibration}

The effective area of the \textit{NICER} XTI is determined primarily using observations of the Crab pulsar and nebula.  The energy-dependent residuals in the fits to the Crab spectrum are typically at the level of $\lesssim2\%$\footnote{See  \url{https://heasarc.gsfc.nasa.gov/docs/heasarc/caldb/nicer/docs/xti/NICER-xti20200722-Release-Notesb.pdf} for further details.}.  The calibration accuracy for the \textit{XMM-Newton} EPIC-MOS and EPIC-pn cameras is determined to better than 3\% and 2\% (at 1$\sigma$), respectively \footnote{See in particular Table 1 in \url{https://xmmweb.esac.esa.int/docs/documents/CAL-TN-0018.pdf}.}. However, in the absence of a suitable absolute calibration source, the uncertainties in the absolute energy-independent effective areas that we use for the \textit{NICER} XTI and the three \textit{XMM-Newton} detectors are estimated to be no more than $\pm 10$\% (e.g., Figure C2 in \citealt{2021arXiv210205666R}; see Section~\ref{sec:relative-normalization} for further discussion).
\bigskip

\section{METHODS}
\label{sec:methods}

Our approach to modeling the X-ray data is largely the same as it was in \citet{2019ApJ...887L..24M}, except that for our analysis of PSR~J0740$+$6620 we used both \textit{NICER} and \textit{XMM-Newton} data, and we use blank-sky observations to estimate the non-source counts in the \textit{XMM-Newton} data.  In Section~\ref{sec:modeling-the-emission} through Section \ref{sec:waveform-models}, we summarize briefly the approaches used in \citet{2019ApJ...887L..24M}, and refer the reader to that paper for details and tests of our procedures.  In Section~\ref{sec:fitting-procedure} we discuss the new aspects of our fitting procedure, which include analyzing multiple data sets and incorporating independent estimates of the background made using the \textit{XMM-Newton} data.  We conclude this section with short discussions of our parameter estimation procedure in Section~\ref{sec:parameter-estimation} (which is the same as was described in \citealt{2019ApJ...887L..24M}), our choice of energy channels in Section~\ref{sec:energy-channels} (which is different from the choice made in \citealt{2019ApJ...887L..24M} because we are using an updated energy calibration for \textit{NICER} and because we also use \textit{XMM-Newton} data in our analysis), and the effects of a possible difference in the absolute calibration of \textit{NICER} and \textit{XMM-Newton} in Section~\ref{sec:relative-normalization}.

\subsection{Modeling the Emission from the Stellar Surface}
\label{sec:modeling-the-emission}

The soft X-ray emission from old, non-accreting pulsars such as PSR~J0740$+$6620 is believed to be generated when particles with large ($\gg 100$) Lorentz factors are produced in pair cascades and deposit their energy deep in the atmosphere of the neutron star (\citealt{1974RvMP...46..815T,2001ApJ...556..987H,2002ApJ...568..862H,2007ApJ...670..668B}; see \citealt{2019ApJ...872..162B} and \citealt{2020A&A...641A..15S} for explorations of the consequences for the beaming pattern of emission from the stellar surface if significant energy is in particles with lower Lorentz factors).  The specific intensity as a function of the angle between the photon propagation direction and the local surface normal depends on the assumed composition, the ionization state, and the strength (and potentially the orientation) of the surface magnetic field at the point of emission.  

We follow \citet{2019ApJ...887L..24M} in assuming that the atmosphere is pure hydrogen and that the magnetic field can be neglected.  In more detail:

\textit{Composition of the upper atmosphere}.---The surface gravity of a neutron star is great enough that the lightest element present is expected to float to the top within seconds to minutes (based on extrapolations of the calculations in \citealt{1980ApJ...235..534A}).  As hydrogen is also the most abundant element in the universe and PSR~J0740$+$6620 likely underwent prolonged accretion to reach its current rotational frequency, it is probable that the atmosphere is pure hydrogen \citep{1992ApJ...399..634B,2019MNRAS.484..974W,2020MNRAS.493.4936W}.  Note that there are some accreting neutron star binaries in which the transferred matter is thought to have little to no hydrogen (such as 4U~1820$-$30; see, e.g., \citealt{2002ApJ...566.1045S}), but such binaries have extremely short orbital periods (e.g., \citealt{1987ApJ...312L..17S} find that the orbital period of 4U~1820$-$30 is just 685 seconds).  Thus the binary is very compact and as a result the hydrogen envelope is believed to have been completely stripped off.  In contrast, the binary containing PSR~J0740$+$6620 has an orbital period of 4.77 days \citep{2020NatAs...4...72C} and the companion could therefore have had a significant hydrogen envelope.  Even if the accreted elements were primarily heavier than hydrogen, spallation might create enough hydrogen to dominate the atmosphere \citep{1992ApJ...384..143B,2019ApJ...887..100R}.

\textit{Ionization fraction}.---The hot spots on PSR~J0030$+$0451 were inferred to have effective temperatures (i.e., the temperature of blackbody that would produce a bolometric photon flux equal to the bolometric photon flux produced by the model atmosphere we are considering, which is not a blackbody) $kT_{\rm eff}>0.1$~keV \citep{2019ApJ...887L..24M,2019ApJ...887L..21R}, which implies that essentially all atmospheric hydrogen is completely ionized.  The fits we find for PSR~J0740$+$6620 have effective temperatures $kT_{\rm eff}\gtorder 0.08$~keV for both spots, which also implies nearly full ionization. However, the atmospheric densities are high enough that some neutral atoms may be present.   We therefore performed fits using the NSX code for model atmospheres assuming full ionization \citep{2001MNRAS.327.1081H} and also with partial ionization (\citealt{2009Natur.462...71H}; see \citealt{2005MNRAS.360..458B} for the relevant opacity tables).  We find no significant difference in the inferred mass and radius between the results obtained using our two different ionization assumptions.  This is consistent with previous work \citep{2013ApJ...776...19L,2015ApJ...808...31M}, which finds that in analyses of phase-channel X-ray data such as is collected using \textit{NICER}, even if the assumptions used in the fit deviate from the properties of the star there is no significant bias in the mass and radius if the fit appears to be statistically good (as measured by, e.g., $\chi^2/{\rm dof}$ or other standard statistics).  Here we report only on runs that allow for the possibility of partial ionization.

\textit{Effect of the magnetic field on the soft X-ray emission from the stellar surface}.---Using Equation~(12) of \citet{2006ApJ...643.1139C}, assuming a magnetic field inclination of $\pi/2$ and $\alpha=0$ for the magnetic field configuration, the period $P=0.00289$~seconds and period derivative ${\dot P}=1.219\times 10^{-20}$ of PSR~J0740$+$6620 \citep{2020NatAs...4...72C} imply that for a centered dipole the polar magnetic field at the surface is $B\approx 3\times 10^8$~G.  The electron cyclotron energy at that field strength is $\approx 3$~eV, which has a small effect on atomic structure \citep{2001RvMP...73..629L,2014PhyU...57..735P} and is $\sim 100$ times lower than the lowest energies we consider in our fits.  The spot configuration resulting from our analysis is close enough to a centered dipole that it is plausible that the effects of the field on radiative transfer are negligible.  For example, in our best fit the two spots are separated by $\approx 2$~radians on the surface.  The chord length is therefore $\approx\sin(1)\approx 0.84$ times smaller than a diameter, so if the magnetic field is a dipole centered between the two spots then the field strength at the surface is $\approx 1/0.84^3\approx 1.7$ times larger than it would be if the field were centered at the center of the star.  Thus in this example the surface magnetic field at the spots would increase only to $\approx 5\times 10^8$~G.  We follow \citet{2019ApJ...887L..24M} in noting that if the field is strong enough to change significantly the spectrum and beaming pattern, then fits assuming a negligible field would likely be statistically poor and/or would yield unreasonable parameter values.  Thus the good quality of our fits could be considered an argument in favor of our assumption that the field can be neglected.  However, this conclusion is not certain and a rigorous resolution of this question would require the construction, and use in fits, of atmospheric models with a fine grid of magnetic field strengths and orientations to the local surface normal.  Such models are challenging to compute at the needed $B\sim 10^9-10^{10}$~G because at those field strengths Coulomb and magnetic effects are comparable to each other.

\subsection{Our Approach to Modeling the Soft X-ray Waveform}
\label{sec:waveform-modeling}

Our modeling of the \textit{NICER} pulse waveform of PSR~J0740$+$6620 builds on the modeling of the waveform of PSR~J0030$+$0451 discussed in \citet{2019ApJ...887L..24M}; see \citet{2019ApJ...887L..26B,2021arXiv210406928B} for detailed descriptions of how we computed pulse waveforms and tested their accuracy.  The waveform of PSR~J0740$+$6620 has two clear peaks and therefore cannot be modeled using a star with a single circular hot spot.  Unlike for PSR~J0030$+$0451, and possibly related to the faintness of PSR J0740+6620, and hence the much smaller number of counts in its pulse waveform, we find that a model with two uniform circular hot spots produces a waveform that adequately fits the data, and that adding complexity to the temperature pattern on the stellar surface does not improve the fit. For example, adding a third hot spot to the model or allowing the spots to be oval rather than circular did not improve the fit.  We therefore fit the \textit{NICER} and \textit{XMM-Newton} data for PSR~J0740$+$6620 using a pulse waveform produced by two uniform-temperature circular spots.

Once the number of spots to be considered and their allowed shapes (e.g., oval or circular) are specified, our spot modeling algorithm automatically explores the full parameter space, to find the configurations with the highest likelihoods (see \citealt{2019ApJ...887L..24M}). The algorithm not only considers configurations in which the spots are disjoint or in contact, but also configurations in which they partially or completely overlap. This allows the algorithm to model approximately spots with temperature gradients, if these are favored by the data, by allowing spots with different temperatures to partially or completely overlap. This is done by assigning a number to each of the spots. For example, for the two-spot model used here, the spots are labeled 1 and 2. Points on the stellar surface that are covered by both spots are assumed to emit with the effective temperature of the lower-numbered spot. This labeling, and the precedence given to the emission by the lower-numbered spot, is not related to any other property of the spot. For example, the effective temperature of spot 1 could be either higher or lower than the effective temperature of spot 2. This approach is highly flexible, and can be generalized to any number of spots (see \citealt{2019ApJ...887L..24M} for details). For example, in addition to the best-fit solution we feature in this paper, our algorithm also found a lower-likelihood solution with a large, crescent-shaped emitting region that it modeled using a spot with a temperature so low that it is essentially dark lying on top of a hotter spot. This solution is disfavored by a Bayes factor greater than 3000 as measured by MultiNest [see Section 3.5], so we have set it aside for the purposes of this paper. 

Section 4 of \citet{2019ApJ...887L..24M} reported the results of an analysis of synthetic \textit{NICER} pulse waveform data constructed assuming two, uniform-temperature, circular hot spots, and separately, an analysis of synthetic data assuming two, uniform-temperature, oval hot spots. In both cases, the values of the model parameters inferred from the analysis was consistent with the values assumed in the constructing the synthetic pulse waveform data. As another test of our analysis algorithm, in Section~\ref{sec:synthetic} of this paper we report a joint analysis of synthetic \textit{NICER} and \textit{XMM-Newton} pulse waveform data sets that were generated using the hot spot model that we found provides a good description of the actual \textit{NICER} and \textit{XMM-Newton} data on PSR~J0740$+$6620. This analysis includes analyses of four separate data sets, each with a realistic background.  

\subsection{Waveform Models}
\label{sec:waveform-models}

The parameters used in the pulse waveform modeling we report here are listed and defined in Table~\ref{tab:wf-primary-parameters}, with their assumed priors.  We assume that the effective temperature is uniform over a given spot.  

Because PSR~J0740$+$6620 is in a binary system, we have additional information on some of the parameters, beyond that provided by the \textit{NICER} and \textit{XMM-Newton} data.  In particular, \citet{2021arXiv210400880F} find that at 68\% credibility the mass is $M=2.08^{+0.072}_{-0.069}~M_\odot$, the distance is $1.136^{+0.174}_{-0.152}$~kpc, and our line of sight makes an angle of $87.5^\circ\pm 0.17^\circ$ to the orbital axis of the system.

We caution that other lines of evidence imply that PSR~J0740$+$6620 has a lower mass.  Prominent among these is that there is an observationally established correlation between the white dwarf mass and the orbital period in white-dwarf binaries \citep{1999A&A...350..928T,2014ApJ...781L..13T}, which would suggest a strong upper limit of $2~M_\odot$ to the mass of PSR~J0740$+$6620.  If the mass is indeed less than our current best estimate, the main effect will be to decrease our estimate of the radius.  Although it is not generally true that the compactness $GM/(Rc^2)$ is measured to better fractional precision than the radius (see Section~4 of \citealt{2021arXiv210406928B}), this is true for PSR~J0740$+$6620 because, as we discuss in more detail below, the most important information about the radius comes from the modulation fraction. A star that is more compact will deflect light rays coming from its surface by a larger angle, allowing a given observer to see more of the surface, which reduces the modulation fraction the observer sees.  As a result, this estimate of the radius of PSR J0740+6620 is roughly proportional to its mass.  Because the fractional uncertainty in the compactness of PSR~J0740$+$6620 is large, even an estimated mass of (for example) $1.8~M_\odot$ would produce a radius posterior that largely overlaps the posterior we report here.

We use the radio observations of \citet{2021arXiv210400880F} to impose a Gaussian prior on the mass of $M=2.08\pm 0.09~M_\odot$, where we have linearly added an estimated systematic error of $0.02~M_\odot$ to the mass estimate from \citet{2021arXiv210400880F} (E. Fonseca, personal communication), and an asymmetric Gaussian prior probability distribution centered at $d = 1.136$~kpc with 68\% cumulative probability points at $d = 0.956$~kpc and $d = 1.336$~kpc, where we have placed the 68\% cumulative probability points 0.03~kpc further from the central value than the quoted statistical uncertainty, to reflect the 0.03~kpc estimated systematic error (E. Fonseca, personal communication).  

Our analysis depends on the inclination of our line of sight to the pulsar's rotational axis.  Millisecond pulsars such as PSR~J0740$+$6620 are believed to be recycled by accreting matter from their companion stars (see, e.g., \citealt{1991PhR...203....1B}). Consequently, the orientation of their rotational axis is thought to be determined by the angular momentum of the matter accreted from the companion; accretion of this matter is expected to gradually align the pulsar's rotational axis with the orbital axis of the system. However, this alignment is not expected to be perfect, and the current rotational axis of the pulsar may therefore be tilted by a few degrees relative to the orbital axis of the binary system. For the analyses reported in this paper, we have adopted a prior on the angle between the observer's line of sight and the pulsars rotation axis that is flat within $5^\circ$ of the best estimate of the system's orbital inclination, which is $87.5^\circ$, and zero outside this range. We find that the inferred mass and radius of PSR~J0740$+$6620 are insensitive to the precise value of the orbital inclination with this range (see the corner plots in the appendices and the discussion in Section~\ref{sec:priors}).

\begin{deluxetable*}{c|l|c}
    \tablecaption{Primary parameters of the pulse waveform models considered in this work.}
\tablewidth{0pt}
\tablehead{
      \colhead{Parameter} & \colhead{Definition} & \colhead{Assumed Prior}
    \label{tab:wf-primary-parameters}
}
\startdata
      \hline
      $c^2R_e/(GM)$ & Inverse of stellar compactness & $3.2-8.0$ \\
      \hline
      $M$ & Gravitational mass & $\exp[-(M-2.08~M_\odot)^2/2(0.09~M_\odot)^2]$ \\
      \hline
      $\theta_{\rm c1}$ & Colatitude of spot 1 center & $0$ to $\pi$ radians \\
      \hline
      $\Delta\theta_1$ & Spot 1 radius & $0-3$ radians \\
      \hline
      $kT_{\rm eff,1}$ & Spot 1 effective temperature & $0.011-0.5$~keV\\
      \hline
      $\Delta\phi_2$ & Spot 2 longitude difference & $0-1$ cycles \\
      \hline
      $\theta_{\rm c2}$ & Colatitude of spot 2 center & $0$ to $\pi$ radians \\
      \hline
      $\Delta\theta_2$ & Spot 2 radius & $0-3$ radians \\
      \hline
      $kT_{\rm eff,2}$ & Spot 2 effective temperature & $0.011-0.5$~keV \\
      \hline
      $\theta_{\rm obs}$ & Observer inclination & $1.44-1.62$ radians \\
      \hline
      $N_H$ & Neutral H column density & $0-20\times 10^{20}$~cm$^{-2}$ \\
      \hline
      $d$ & Distance & $\exp[-(d-1.136~{\rm kpc})^2/2(0.2~{\rm kpc})^2]$, $d\geq 1.136~{\rm kpc}$ \\
          & & $\exp[-(d-1.136~{\rm kpc})^2/2(0.18~{\rm kpc})^2]$, $d\leq 1.136~{\rm kpc}$ \\
      \hline
      $A_{\rm XMM}$ & \textit{XMM-Newton} effective area\\
          & divided by nominal area & $0.9-1.1$\\
      \hline 
\enddata
\tablecomments{Except where noted, the prior is flat over the given range.}
\end{deluxetable*}

\subsection{Our Waveform Modeling Procedure}
\label{sec:fitting-procedure} 

For PSR~J0030$+$0451 we had only \textit{NICER} data to analyze.  In contrast, for PSR~J0740$+$6620 we also have \textit{XMM-Newton} imaging data from pn, MOS1, and MOS2.  We therefore need to model all four data sets jointly.  Once we have full models including backgrounds for all four data sets, then the log likelihood of the data given the model is simply the sum of the log likelihoods of each of the individual data sets.  Thus
\begin{equation}
\ln{\cal L}_{\rm tot}=\ln{\cal L}_{\rm NICER}+\ln{\cal L}_{\rm XMM-pn}+\ln{\cal L}_{\rm XMM-MOS1}+\ln{\cal L}_{\rm XMM-MOS2}\; .
\end{equation}
In writing the total log likelihood this way, we are assuming that the data sets are uncorrelated with each other.  This assumption is clearly justified for the independent \textit{NICER} and \textit{XMM-Newton} data.  It is also reasonable for the three \textit{XMM-Newton} data sets because the pn, MOS1, and MOS2 cameras are all separately mounted and read out on \textit{XMM-Newton}; there is no plausible mechanism that would link counts in one detector with counts in another detector. 

For each individual data set we compute the log likelihood by summing the log Poisson likelihood of the data given the model in all of the phase bins and energy channels, neglecting the log factorial term that is common to all models.  That is, if we have broken the data into bins $\phi_i$ in rotational phase, and energy channels $E_j$, and in each phase-channel bin the model predicts $m_{ij}$ counts (a positive real number) and $d_{ij}$ counts are observed (a nonnegative integer) then the Poisson likelihood of the data given the model in that phase-channel bin is $[m_{ij}^{d_{ij}}/d_{ij}!]e^{-m_{ij}}$.  However, the factor $1/d_{ij}!$ is common to all models and hence can be neglected in both parameter estimation and model comparison.  The log likelihood we use is therefore
\begin{equation}
\ln{\cal L}=\sum_{\phi_i}\sum_{E_j}\left(d_{ij}\ln m_{ij}-m_{ij}\right)\; ,
\end{equation}
where there is an implied sum over all four instruments.  The \textit{NICER} data have 32 phase bins.  In contrast, for each of the three \textit{XMM-Newton} data sets, we have only one bin in rotational phase because the timing resolution is insufficient to divide the data further.

We now discuss the two components of the calculation of $m_{ij}$ for each phase-channel bin in each data set: the computation of the model counts from the spots for each of the four instruments, which is straightforward, and the incorporation of phase-independent background counts, which is more involved and requires different treatments for the \textit{NICER} and for the \textit{XMM-Newton} data.  
\bigskip
\ 
\bigskip

\subsubsection{Calculation of the model counts from the spots}

Suppose that we are considering a particular parameter combination, i.e., we have some combination of $M$, $R_e$, $\theta_{\rm obs}, \ldots$ to assess.  Given the specified observation time, this gives us a spots-only waveform that is unique up to an arbitrary definition of the time corresponding to phase 0.  As described in Section~3.4 of \citet{2019ApJ...887L..24M}, once we have a candidate waveform that includes both the contribution from the spots and the contribution from the phase-independent background, we marginalize over the overall phase by fitting a Gaussian to the likelihood as a function of phase (this is equivalent to simply fitting a parabola to the log likelihood near its peak).  We only need to carry out this process for the \textit{NICER} data, because the \textit{XMM-Newton} data have only one phase.

\subsubsection{Inclusion of unmodulated background}
\label{sec:background} 

By ``background" we mean any X-ray counts that are not contributed by the spots.  This could include space weather, optical loading, resolved or unresolved background or foreground sources, the instruments themselves, or in the case of a binary such as PSR~J0740$+$6620, the companion star (e.g., interactions of the neutron star particle wind with the companion star could produce X-rays).  What these have in common is that those X-rays are not modulated at the rotational frequency of the star and hence with enough exposure time they contribute equally at all rotational phases.  

\citet{2019ApJ...887L..24M} therefore treated all non-stellar X-rays with a single approach, which we adopt here for the \textit{NICER} data: for each energy channel independently we allow there to be a phase-independent component of any magnitude.  We therefore do not assume any particular spectral form for the non-stellar emission, but in practice this results in an estimated background that varies smoothly with energy channel.  The assumption of independence between channels means that we can fit a Gaussian to the likelihood of each energy channel as a function of added background, and can then analytically integrate the likelihood to marginalize over the background.  See \citet{2019ApJ...887L..24M} for more details.  We also note (see Section~\ref{sec:estimating-J0740-mass+radius} for a more thorough discussion) that future analyses may be able to incorporate estimates of the \textit{NICER} background rather than letting the background be completely free.

The \textit{XMM-Newton} data must be treated differently, because there is only one phase bin, in contrast to the 32 rotational phase bins that we use for the \textit{NICER} data.  To understand why, suppose that we treat the \textit{XMM-Newton} background, for any instrument, in the same way that we treat the \textit{NICER} background: each energy channel (with its single phase) can have added to it any number of background counts.  Then the best model for the \textit{XMM-Newton} data would have zero counts from the spots, and the background would equal the observed number of counts.  This would automatically maximize the log likelihood.  The \textit{NICER} data are not driven to this pathological fit because there is clear modulation in the data; a phase-independent, background-only fit cannot reproduce this modulation.  

Therefore, we instead assume that the blank-sky background is a Poisson realization of the full \textit{XMM-Newton} background.  Each of the \textit{XMM-Newton} instruments has had, throughout the course of the mission, significant exposure to portions of the sky with no known X-ray sources (see Section~\ref{sec:observations}).  The total blank-sky count rate in the \textit{XMM-Newton} energy channels of interest is $\sim 20-30$\% of the count rate in our \textit{XMM-Newton} observations of PSR~J0740$+$6620, so incorporation of this background is expected to make a meaningful difference to our analysis.  We note that there are other possible contributions than the blank sky to the phase-independent background.  For example, as mentioned above, the interaction of the pulsar wind with the binary companion could produce X-rays.  Because the total count rate is fixed, if the background is larger then the estimated total count rate from the neutron star is reduced.  As a result, the modulation fraction is increased because the modulated count rate is fixed by the \textit{NICER} observations.  In turn, this would increase the inferred radius for PSR~J0740$+$6620.  In our treatment we effectively assume that these other contributions can be neglected.

Even with hundreds of thousands of seconds of total exposure time, the total number of counts from the blank sky observations is too small in many of the energy channels to use Gaussian statistics.  Therefore we need to use Poisson statistics.  Our approach to the pn, MOS1, and MOS2 backgrounds is as follows.

Consider some particular \textit{XMM-Newton} energy channel $n$ for one of the three instruments.  Suppose that the blank-sky observation had duration $T_{\rm back}$ and yielded $b_{\rm back}$ counts.  If the prior on the background counts is flat, then the Poisson probability distribution for the time-averaged number of counts $b_{\rm avg}$ for an interval $T_{\rm back}$ is
\begin{equation}
P(b_{\rm avg})db_{\rm avg}=(1/b_{\rm back}!)b_{\rm avg}^{b_{\rm back}}\exp(-b_{\rm avg})db_{\rm avg}\; .
\label{eq:avgrate}
\end{equation}

Now suppose that for the \textit{XMM-Newton} instrument under consideration the PSR~J0740$+$6620 observation has a duration of $T_{\rm obs}$.  Then a given $b_{\rm avg}$ implies an expected number of counts $b_{\rm obs}=b_{\rm avg}(T_{\rm obs}/T_{\rm back})$ counts in that observation.  Then because $b_{\rm avg}=(T_{\rm back}/T_{\rm obs})b_{\rm obs}$, Equation~(\ref{eq:avgrate}) implies that the normalized probability distribution for $b_{\rm obs}$ is
\begin{equation}
P(b_{\rm obs})db_{\rm obs}=(1/b_{\rm back}!)[b_{\rm obs}(T_{\rm back}/T_{\rm obs})]^{b_{\rm back}}\exp[-b_{\rm obs}(T_{\rm back}/T_{\rm obs})](T_{\rm back}/T_{\rm obs})db_{\rm obs}\; .
\end{equation}

If the energy channel had $d$ observed counts in our PSR~J0740$+$6620 observation, then the probability of obtaining that in a model with $m$ expected counts is
\begin{equation}
p(d|m)=m^d/d!~\exp(-m)\; .
\end{equation}
If in our model we have $s$ expected counts from the spots, then given a background $b$ we expect a total number of counts $m=s+b$.  Setting $b=b_{\rm obs}$ as above, we find that the final likelihood of the data $d$ given the spot counts $s$ and the distribution $P(b_{\rm obs})db_{\rm obs}$ is
\begin{equation}
   p(d|s,{\rm back})=\int_0^\infty [(s+b_{\rm obs})^d/d!]\exp[-(s+b_{\rm obs})] P(b_{\rm obs}) db_{\rm obs}\; .
\end{equation}

We find that we achieve sufficient precision for this integral by (1) setting a scale for $b_{\rm obs}$ that is $T_{\rm obs}/T_{\rm back}$ if $b_{\rm back}=0$, or $b_{\rm back}(T_{\rm obs}/T_{\rm back}$) otherwise, and (2) performing a Simpson's rule integration from 0 to 6.8 times that scale, in intervals of 0.1 times that scale.

\subsection{Parameter Estimation and Model Evaluation}
\label{sec:parameter-estimation}

As in \citet{2019ApJ...887L..24M}, we began our sampling of parameter space using the publicly available nested sampler MultiNest \citep{2009MNRAS.398.1601F}, typically with 1000 live points, a target efficiency of 0.01 (with variable efficiency), and a tolerance of 0.1.  The inverse of the efficiency parameter determines the factor by which the bounding hyperellipsoids constructed by MultiNest are expanded before samples are drawn from within them (see Section 5.2 of \citealt{2009MNRAS.398.1601F}).  Lower efficiency means more thorough sampling, and better accommodation of isolikelihood surfaces that are not well-described by ellipsoids.  We chose an efficiency of 0.01, which is more than an order of magnitude below the standard recommendation, because we found in early tests that when we fixed the number of live points to 1000 and used efficiencies of 0.1 and 0.03 the $\pm 1\sigma$ credible regions for the radius were, respectively, 45\% and 22\% narrower than for our efficiency=0.01 runs, and thus that the sampling at these higher efficiencies with 1000 live points was incomplete.   

MultiNest is designed to estimate the Bayesian evidence for a model, and can also provide a quick characterization of parameter space.  However, the purpose of nested samplers such as MultiNest is to compute the evidence; convergence of that calculation does not guarantee convergence of the posteriors of the parameters.  Studies have shown that nested samplers of this type can give incorrectly small credible regions if isolikelihood surfaces cannot be well-represented by overlapping hyperellipsoids \citep{2014SaC...2014.1B,2020AJ....159...73N,2021arXiv210109675B}.  This appears be the case for our PSR~J0740$+$6620 parameter estimation; we found that the parameter space was not completely explored even at the lowest efficiencies and largest number of live points that we employed.  As one example, when we used MultiNest to analyze just the \textit{NICER} data with 1000 live points, a target efficiency of 0.01, and a tolerance of 0.1, the $\pm 1$ standard deviation range in the radius posterior had 67\% of the width we obtained using parallel-tempered emcee (see the next paragraph).  When we used 3000 live points, a target efficiency of 0.01, and a tolerance of 0.1 the width increased, but only to 79\% of the emcee width.

We therefore use a kernel density estimate of the MultiNest posteriors to produce the initial positions of the walkers, which we find expedites convergence, especially for multi-modal posteriors,  in the parallel-tempered Markov chain Monte Carlo (MCMC) sampler PT-emcee that is included in the publicly available emcee package, version 2.2.1 \citep{2013PASP..125..306F}.  Because MCMC samplers are designed to satisfy detailed balance, there is a theoretical expectation that the samples will represent the posterior.  For the PT-emcee runs we had 1000 walkers per parallel-tempering temperature rung; exploratory runs with 2000 walkers per rung produced results that were consistent with what we report here.  We started with five temperature rungs, but found that the number of rungs was not critical and reduced the number to two later in the sampling.

\subsection{Choice of Energy Channels}
\label{sec:energy-channels}

In their analysis of the PSR~J0030$+$0451 data, \citet{2019ApJ...887L..24M} used only \textit{NICER} energy channels 40 and above, because they found that inclusion of lower-energy channels biased the inferred distance.  For our analysis of the PSR~J0740$+$6620 data we use the updated July 2020 \textit{NICER} response (see https://heasarc.gsfc.nasa.gov/docs/heasarc/caldb/nicer/).  With this response there does not appear to be any bias in distance or other quantities when we use data down to energy channel 30.  Analysis of the \textit{NICER} data showed that the detection significance is optimized when we use channels up to 118 (see Section~\ref{sec:data-used}), but there is also an excess of counts relative to the background up to channel $\sim 150$.  We used a range between these (channels 30 through 123 inclusive), but exploratory runs including channels up to 150 did not produce palpably different results.  Given that the width of the \textit{NICER} channels is $0.001$~keV, our energy channel choice corresponds approximately to photon energies between $0.3$~keV and $1.23$~keV.  For the \textit{XMM-Newton} data, we used channels 57$-$299 inclusive for pn, channels 20$-$99 inclusive for MOS1, and channels 20$-$99 inclusive for MOS2.
\bigskip

\subsection{Relative Normalization of \textit{NICER} and \textit{XMM-Newton}}
\label{sec:relative-normalization}

In addition to the uncertainty in the energy-dependent response of an instrument, there is inevitably some uncertainty in the overall response.  That is, the overall normalization of the response is not known perfectly.  A conservative upper limit on the fractional uncertainty of the relative normalization between \textit{NICER} and the three \textit{XMM-Newton} instruments is $\sim 10$\%\footnote{For the cross-calibration as of December 2018 see the lower panel of Figure C2 in \citealt{2021arXiv210205666R}.  For the cross-calibration as of November 2020 see slide 7 of http://iachec.org/wp-content/presentations/2020/NICER-CrossCal-IACHEC-Markwardt-2020b.pdf and note in particular that in the energy range relevant to our analyses, $0.3-1$~keV, the flux inferred from 3C~273 agrees between \textit{NICER} and \textit{XMM-Newton} to much better than 10\%.}.  The three \textit{XMM-Newton} instruments are thought to have the same overall normalization to $\sim 2-3$\%, which are differences small enough to be neglected.  Errors in the overall normalization of \textit{NICER} are absorbed in the $>15$\% half-width of our prior on the distance.  However, there is still a question of whether the overall normalizations of \textit{NICER} and \textit{XMM-Newton} could affect our fits.

To explore this, we report in Section~\ref{sec:estimating-J0740-mass+radius} an additional fit in which we add one more parameter: $A_{\rm XMM}$, which is a single number that multiplies the ARFs of all three \textit{XMM-Newton} instruments that we consider.  We explore the consequences of allowing $A_{\rm XMM}$ to be between $0.9$ and $1.1$, i.e., the largest possible range.  We therefore account for a possible inaccuracy in the ARF of \textit{NICER} by the freedom in the fit to the distance to the source, and for an additional independent inaccuracy of the ARF of the \textit{XMM-Newton} instruments using $A_{\rm XMM}$.  We find that introduction of the parameter $A_{\rm XMM}$ does not produce a significant change in the posteriors of any of our parameters (for example, the radius limits are changed by at most 0.1~km) and the value of $A_{\rm XMM}$ that we infer is consistent with 1.  

\section{RESULTS OF THE ANALYSIS OF THE \textit{NICER} and \textit{XMM-Newton} PULSE WAVEFORM DATA}
\label{sec:fits-to-data}

Our joint fit of our two-circular-spot model to the \textit{NICER} and \textit{XMM-Newton} data yields an equatorial circumferential radius of $R_e=13.7^{+2.6}_{-1.5}$~km at 68\% credibility for PSR~J0740$+$6620.  In this section we describe how we reach this conclusion.  In Section~\ref{sec:synthetic} we demonstrate that our method produces statistically expected results when we fit our model to a synthetic data set that is drawn from a good fit to the PSR~J0740$+$6620 data including backgrounds, and is therefore realistic.  In Section~\ref{sec:estimating-J0740-mass+radius} we present our fits to just the \textit{NICER} data on PSR~J0740$+$6620, and to the \textit{NICER} and \textit{XMM-Newton} data combined.  We show that although the radius posteriors in the two analyses have significant overlap, inclusion of the \textit{XMM-Newton} data favors larger radii.  In Section~\ref{sec:convergence} we show that our sampling of the parameter space has converged.  In Section~\ref{sec:priors} we address the effects of priors, and show that other choices for the upper limit of the radius and the allowed range of the observer inclination have little effect on the $-1\sigma$ radius, which is particularly important in EOS inference (see Section~\ref{sec:EOS}).  In Section~\ref{sec:adequacy} we address the adequacy of our model.  We show that our model provides acceptable fits to the phase-channel \textit{NICER} data, the bolometric \textit{NICER} data, and the \textit{XMM-Newton} data from all three cameras.  There is therefore no indication that our model for the PSR~J0740$+$6620 data is deficient.  Finally, in section~\ref{sec:differences} we note some differences between our analysis method and that of \citet{2021..............R}, and suggest which of those differences could be important in producing different radius posteriors between the two groups.

In this section consider values of $R_{\rm eq}c^2/(GM)$ up to 8.0, which corresponds to a radius $\sim 24$~km at $M\sim 2~M_\odot$.  We find that the radius posterior does not have significant probability near this boundary, and thus that this boundary does not bias our radius estimates.  When we explore the implications of results for the EOS in Section~\ref{sec:EOS}, we include information from nuclear theory and previous measurements.

\subsection{Fits to Synthetic \textit{NICER} and \textit{XMM-Newton} Data}
\label{sec:synthetic}

Joint analyses of \textit{NICER} and \textit{XMM-Newton} data have not previously been performed in the context of neutron star mass and radius estimation.  It is therefore important to demonstrate that our method yields the expected results when similar synthetic data are analyzed, because for synthetic data we know the parameter values used to construct the data and can therefore judge the quality of the fit.

To generate the data we started from a good model of the PSR~J0740$+$6620 \textit{NICER} and \textit{XMM-Newton} data, including all backgrounds.  We then produced the synthetic data by drawing integer numbers of counts from our (in general non-integer) model expectation in each phase-channel bin for \textit{NICER}, and in each energy channel for each of the three \textit{XMM-Newton} cameras.  We emphasize that this leads to synthetic data with fully realistic backgrounds; previous synthetic data analyses (e.g., \citealt{2013ApJ...776...19L,2015ApJ...808...31M,2019ApJ...887L..24M,2021arXiv210406928B}) have instead typically used power-law backgrounds.  This is therefore the most realistic synthetic data test that has been performed in the radius estimation context.

\begin{deluxetable*}{c|c|c|c}
    \tablecaption{Verification of Fits to Synthetic \textit{NICER} and \textit{XMM-Newton} Data}
\tablewidth{0pt}
\tablehead{
      \colhead{Sampler} & \colhead{$\pm 1\sigma$ (68.3\%)} & \colhead{$\pm 2\sigma$ (95.4\%)} & \colhead{$\pm 3 \sigma$ (99.7\%) }
    \label{tab:synth}
}
\startdata
      \hline
      PT-emcee & 6 & 11 & 12\\
      \hline
      MultiNest  & 4 & 7 & 9\\
      (Nlive=1000, efficiency=0.01) & & & \\
      \hline 
      MultiNest  & 4 & 8 & 11\\
      (Nlive=2000, efficiency=0.01) & & & \\
      \hline
\enddata
\tablecomments{Results of fits to synthetic J0740-like \textit{NICER} and \textit{XMM-Newton} data using parallel-tempered emcee and two different settings of MultiNest.  The right three columns show the number of parameters, out of 12, for which the values assumed in constructing the synthetic data are within the $\pm 1\sigma$, $\pm 2\sigma$, and $\pm 3\sigma$ credible regions.  The PT-emcee results are consistent with statistical expectations (see Appendix~\ref{sec:corner-plot-Synthetic} for the full posteriors).  In contrast, the credible regions for both MultiNest runs are too narrow, although with more live points the sampling is more thorough.}
\end{deluxetable*}

The results of our analysis of the synthetic data are summarized in Table~\ref{tab:synth}, and more detail on our PT-emcee analysis is shown in Table~\ref{tab:crediblesynthetic} and Figure~\ref{fig:full-posteriors-for-synthetic} in Appendix~\ref{sec:corner-plot-Synthetic}.  We see that the PT-emcee sampling yields results that are entirely consistent with statistical expectations.  In contrast, MultiNest sampling with Nlive=1000 and an efficiency of 0.01 is not adequate; for example, 3 of the 12 parameters have $\pm 3\sigma$ credible regions that exclude the parameter value used to construct the data.  The MultiNest sampling with Nlive=2000 and an efficiency of 0.01 is better but the credible regions are still too narrow: 1 of the 12 true parameter values is excluded at $>3\sigma$, and 4 of the 12 are excluded at $>2\sigma$.  It is possible that, with more live points and/or lower efficiency, MultiNest analysis of our synthetic data would lead to the statistically expected outcomes, but results such as these form part of our motivation to use PT-emcee to determine posterior probabilities.

In summary, even with realistic backgrounds and even when analyzing jointly the \textit{NICER} and \textit{XMM-Newton} synthetic data, our analysis method yields statistically acceptable results.

\subsection{The Mass and Radius of PSR J0740$+$6620}
\label{sec:estimating-J0740-mass+radius}

\begin{figure*}[ht!]
          \includegraphics[width=0.49\textwidth]{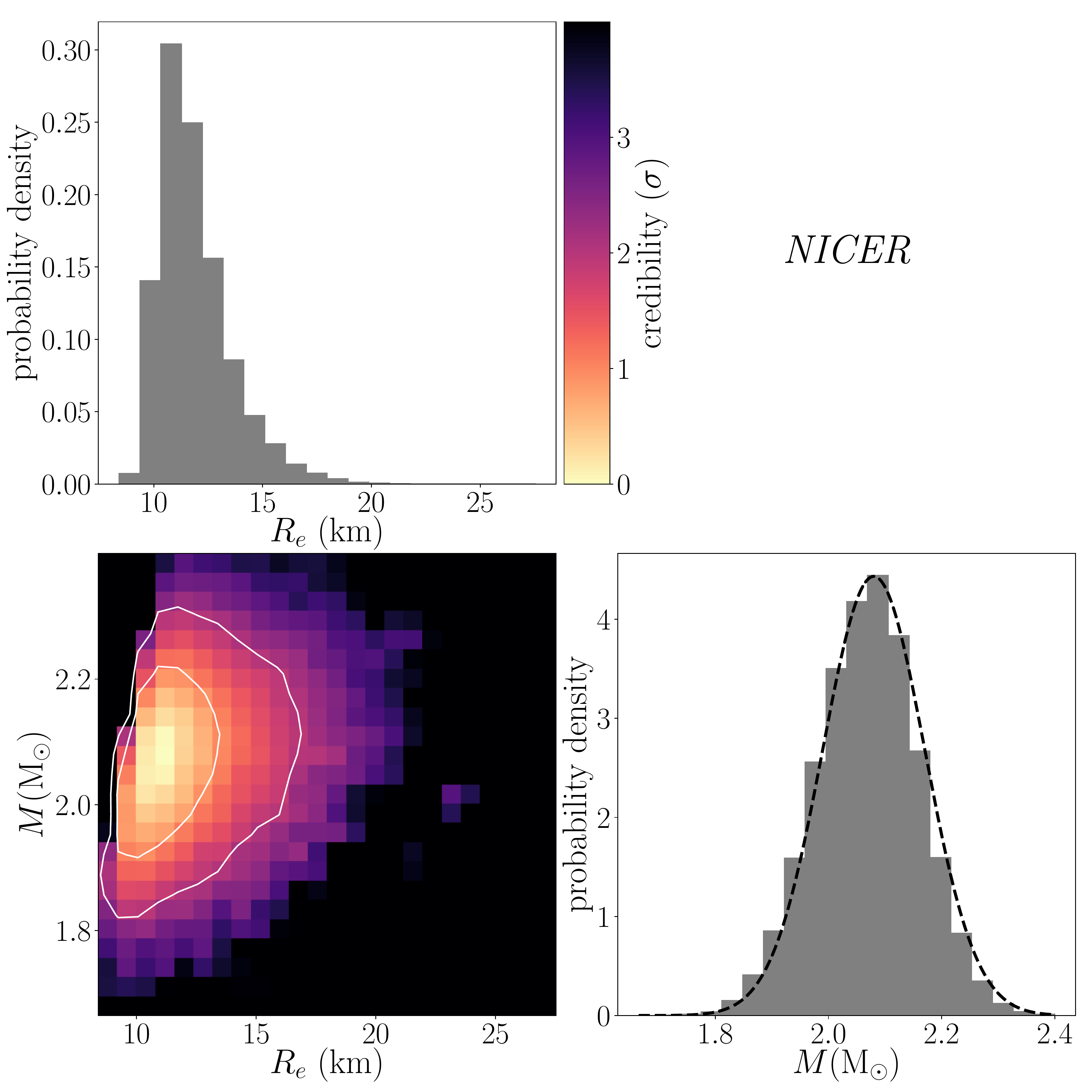}
          \includegraphics[width=0.49\textwidth]{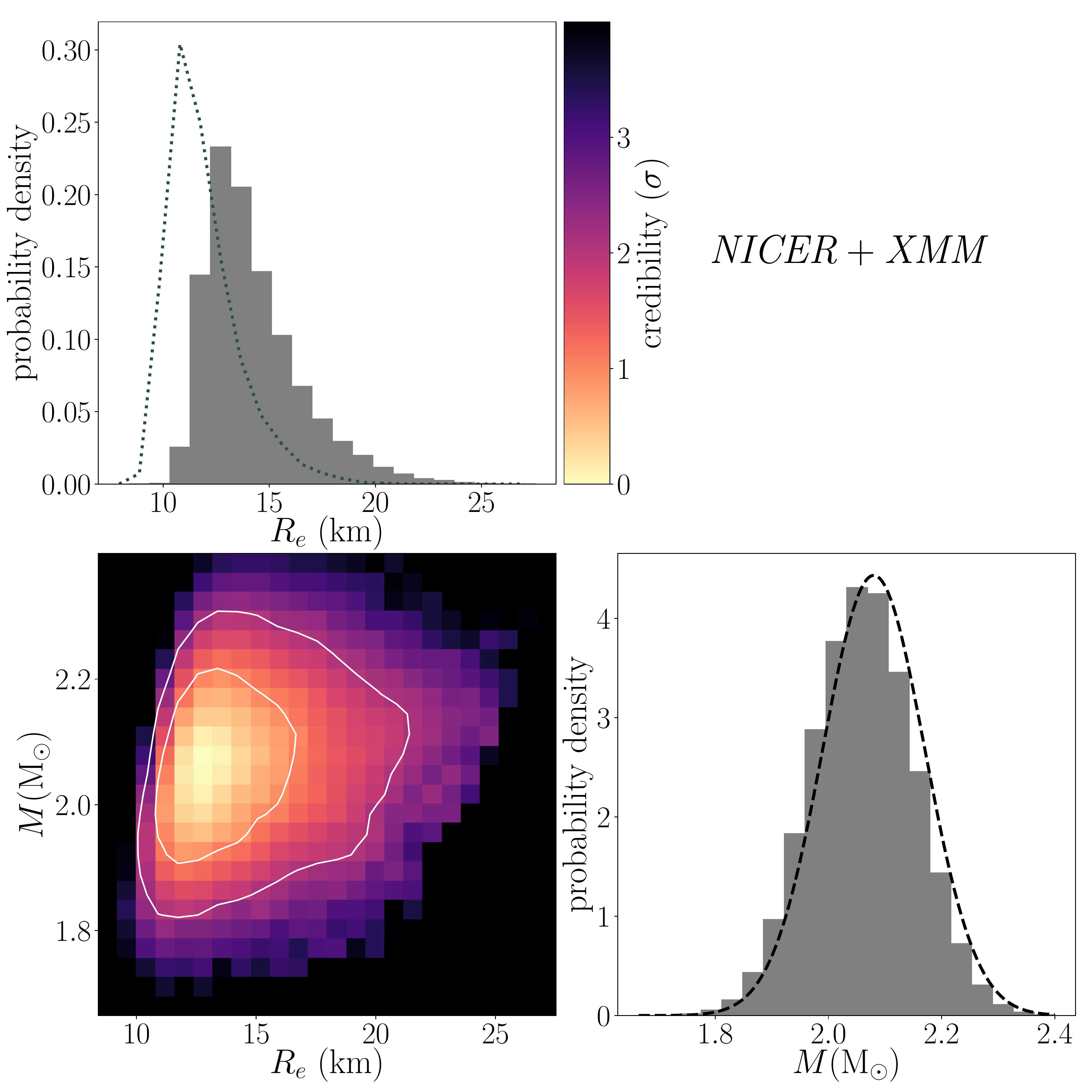}
\caption{Comparison of the $M-R_e$ corner plots obtained using two circular spots, fit to only the \textit{NICER} data (left panels) and to both the \textit{NICER} and the \textit{XMM-Newton} data with the nominal \textit{XMM-Newton} calibration (right panels).  All of the results that we present use pure hydrogen model atmospheres that allow for the possibility of partial ionization.  The mass prior is represented by the dashed line in each of the one-dimensional mass posterior plots.  In the mass-radius plots, brighter colors indicate higher posterior probability densities; the inner contour in these plots contains 68.3\% of the posterior probability whereas the outer contour contains 95.4\%.  In the \textit{NICER}+\textit{XMM-Newton} one-dimensional radius posterior plot, the dotted line shows the posterior obtained using only the \textit{NICER} data.  The two radius posteriors have substantial overlap, but the effect of including the \textit{XMM-Newton} data is to increase the favored radius by roughly 2~km.}
\label{fig:MR-comparison}
\end{figure*}

\begin{figure*}[ht!]
  \begin{minipage}[c]{0.6\textwidth}
    \includegraphics[width=\textwidth]{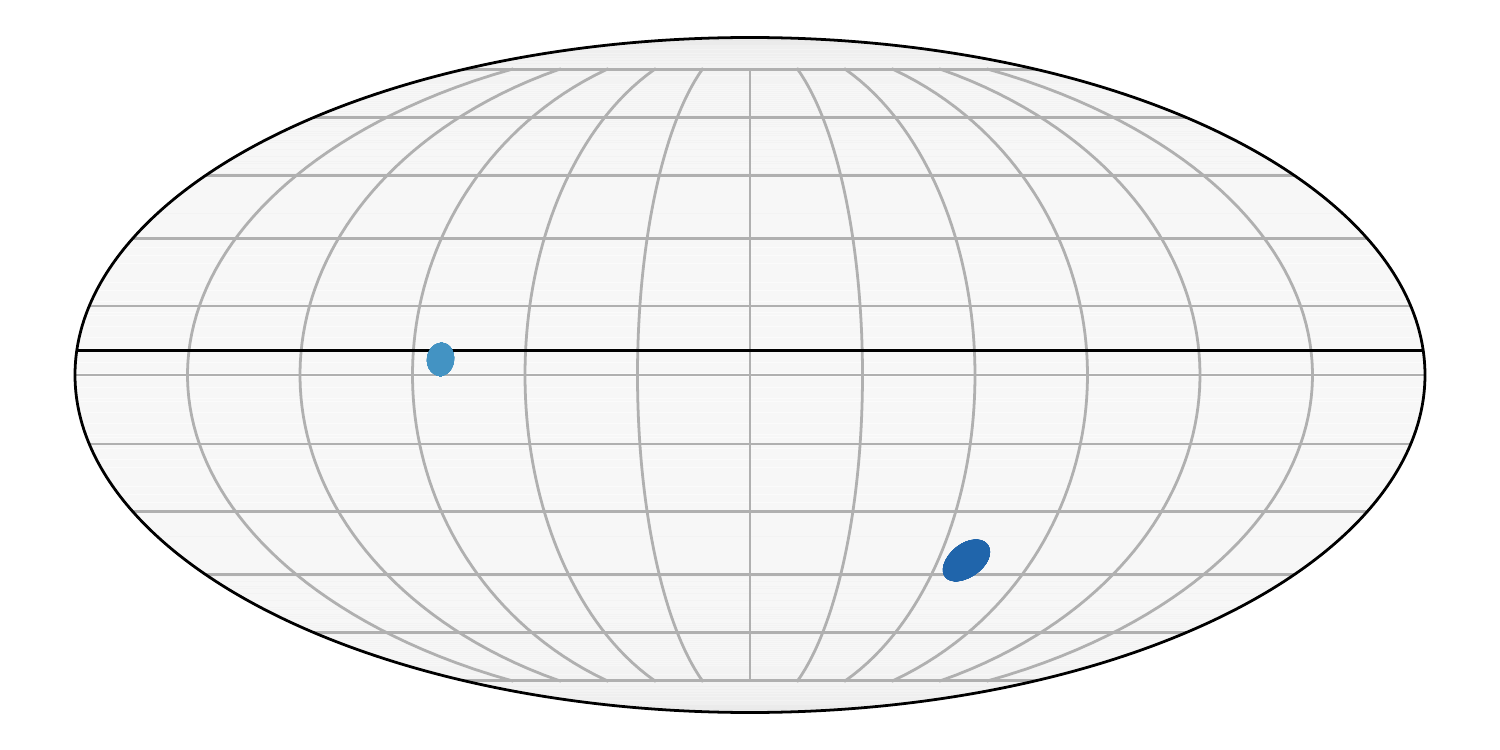}
  \end{minipage}\hfill
  \begin{minipage}[c]{0.35\textwidth}
    \caption{Spot locations for a representative fit, in an equal-area Mollweide projection.  The horizontal bar indicates the colatitude of the observer.  Because we observe from close to the rotational equator, and because the spots do not overlap, there is an approximate four-fold degeneracy in solutions.}       
\label{fig:spot-locations}
  \end{minipage}
\end{figure*}

In this section we present fits to (1)~just the \textit{NICER} data (see Table~\ref{tab:credibleNICERonly} in Appendix~\ref{sec:corner-plot-NICERonly} for the best fit and uncertainties, and the left hand panels of Figure~\ref{fig:MR-comparison} for the mass and radius posteriors), (2)~the \textit{NICER} data and the data from the three \textit{XMM-Newton} instruments, assuming the nominal calibration for \textit{XMM-Newton} (see Table~\ref{tab:credibleNICERXMM} in Appendix~\ref{sec:corner-plot-NICERXMM} for the best fit and uncertainties, and the right hand panels of Figure~\ref{fig:MR-comparison} for the mass and radius posteriors), and (3)~the \textit{NICER} data and the data from the three \textit{XMM-Newton} instruments, with an additional energy-independent factor $0.9<A_{\rm XMM}<1.1$ that multiplies the effective area of the \textit{XMM-Newton} instruments (see Table~\ref{tab:credibleNICERXMMnorm} in Appendix~\ref{sec:corner-plot-NICERXMMcalib} for the best fit and uncertainties).  The appendices have the full corner plots for each of the three fits.  In all cases we use atmosphere tables for pure hydrogen that accommodate partial ionization.

Figure~\ref{fig:spot-locations} shows a representative spot pattern.  We note that because the observer inclination is very close to the rotational equator, there is a near-symmetry between spots in the northern hemisphere (which we define to be the hemisphere of the observer) and the southern hemisphere.  In addition, because the spots do not overlap, their labeling (i.e., which is spot 1 and which is spot 2) can be swapped without affecting the solution, unlike what would be the case if the spots overlapped.  Thus there is an approximate four-fold degeneracy in the solution.

The inclusion of \textit{XMM-Newton} data increases the radius estimates: for example, with the \textit{NICER} data alone the $\pm 1\sigma$ range in radius is 10.382~km to 13.380~km, whereas when we add the \textit{XMM-Newton} data the range becomes 12.209~km to 16.326~km.  This is because the prime source of information about the compactness is the modulation fraction (see \citealt{2016ApJ...822...27M}).  The \textit{NICER} data are dominated by background counts, which means that from that data set alone it could be that the modulation fraction is low.  If other parameters such as the observer inclination and spot locations are fixed, then because light deflection increases with increasing compactness $GM/(R_ec^2)$, for a more compact star the light from spots is spread out more and thus the modulation fraction is reduced.  As a result, the \textit{NICER} data alone can be fit with very compact stars.  But when the \textit{XMM-Newton} data are included, we have a measure of the total flux from the star.  This total flux is significantly less than the total flux in the \textit{NICER} data, because the \textit{NICER} data include significant background.  Thus, the modulation fraction is higher than it appears to be from  the \textit{NICER} data alone, and hence the compactness of the star must  be lower. Because the mass is known with some precision, a lower compactness implies a larger radius.  

We also note that, for the same reason, if reliable information about the non-spot \textit{NICER} flux were included in the \textit{NICER} data then the inferred radius would increase.  That is, any information that places an upper limit on the flux from the spots, whether it comes from observations of the source (in our case, using \textit{XMM-Newton}) or a model of some part of the the non-spot flux (which we would obtain with a background model), increases the inferred modulation fraction and thus increases the inferred radius.  Hence analysis of only the \textit{NICER} data on PSR~J0740$+$6620, if augmented with \textit{NICER} background data that (critically) have reliable estimates of the field-to-field variance as well as statistical uncertainties, is likely to result in an inferred radius that is close to the radius inferred when we also use \textit{XMM-Newton} data.  The 3C50 \textit{NICER} background model of Remillard et al. (2021, submitted; see https://heasarc.gsfc.nasa.gov/docs/nicer/tools/nicer\_bkg\_est\_tools.html) is a good step in this direction, but the variance between the background in different sky pointings is not well understood.

We stress that despite the clear shift of the radius posterior distribution to larger radii when the \textit{XMM-Newton} data are included, this distribution is still broadly consistent with the distribution obtained using the \textit{NICER} data alone, even without background information. One indication of this is that there is a substantial overlap between the two $\pm 1\sigma$ ranges.  To quantify this overlap we can compute, e.g., the Bhattacharyya coefficient \citep{1943BCMS..35...99B}
\begin{equation}
BC(p,q)=\int\sqrt{p(x)q(x)}dx
\end{equation}
for two normalized probability distributions $p(x)$ and $q(x)$.  $BC=1$ for identical distributions, whereas $BC=0$ for distributions with zero overlap.  In our case, $BC=0.77$ when $p$ is the radius posterior using only the \textit{NICER} data and $q$ is the radius posterior when the \textit{XMM-Newton} data are also included, and when we represent those distributions using histograms with bin widths of 0.1~km.  For context, $BC=0.77$ for two unit-variance Gaussians whose centers are separated by roughly 1.5 times their standard deviations.  This indicates that there is a significant overlap between the two distributions.

\begin{figure*}[ht!]
\vspace{-1.8truein}
          \includegraphics[width=\textwidth]{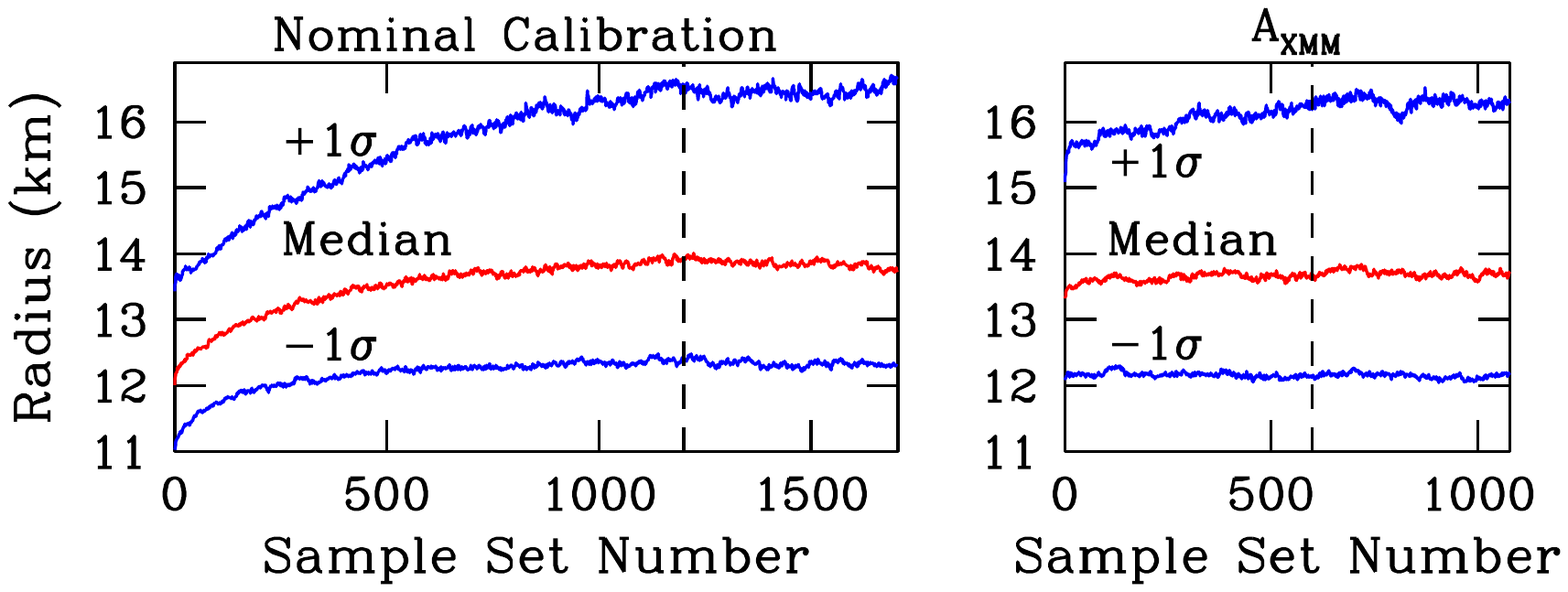}
\vspace{-6.2truein}
\caption{Evolution of the $-1\sigma$, median, and $+1\sigma$ values of the radius posterior as our PT-emcee sampling of the \textit{NICER}+\textit{XMM-Newton} data progressed.  Each point in the plots indicates the estimated values of these three radii given by a single set of 20,000 contiguous, non-overlapping samples.  The number of sets included in each estimate is shown by the numbers on the horizontal axis.  Left: Evolution of these three radii as sampling proceeds, starting from the final posterior distribution provided by MultiNest, and assuming the nominal ratio of the \textit{NICER} overall effective area relative to that of \textit{XMM-Newton}.  As sampling proceeded, all three radii increased rapidly and then converged to values that are larger, with a broader spread, than was obtained from the MultiNest sampling.  The sample sets to the right of the dashed vertical line were used to construct the final posterior distribution from this run (a total of $10^7$ samples), which is what we used in our parameter estimation.  Right: Evolution of the three radii as sampling proceeded, starting with a kernel density estimate of the posterior distribution obtained at the end of the previous run, but now including the overall effective area $A_{\rm XMM}$ of \textit{XMM-Newton} as a parameter in the fit. The change in the treatment of the effective area produced a small shift in the values of these three radii relative to their final values in the left panel.  As sampling progressed, the three radii rapidly approached the values they had at the end of the run shown in the left panel.  The $10^7$ samples to the right of the dashed vertical line were used to construct the posterior distribution for this run.  See the main text for more details.}       
\label{fig:convergence}
\end{figure*}

\begin{figure*}[ht!]
  \begin{minipage}[c]{0.43\textwidth}
          \includegraphics[width=\textwidth]{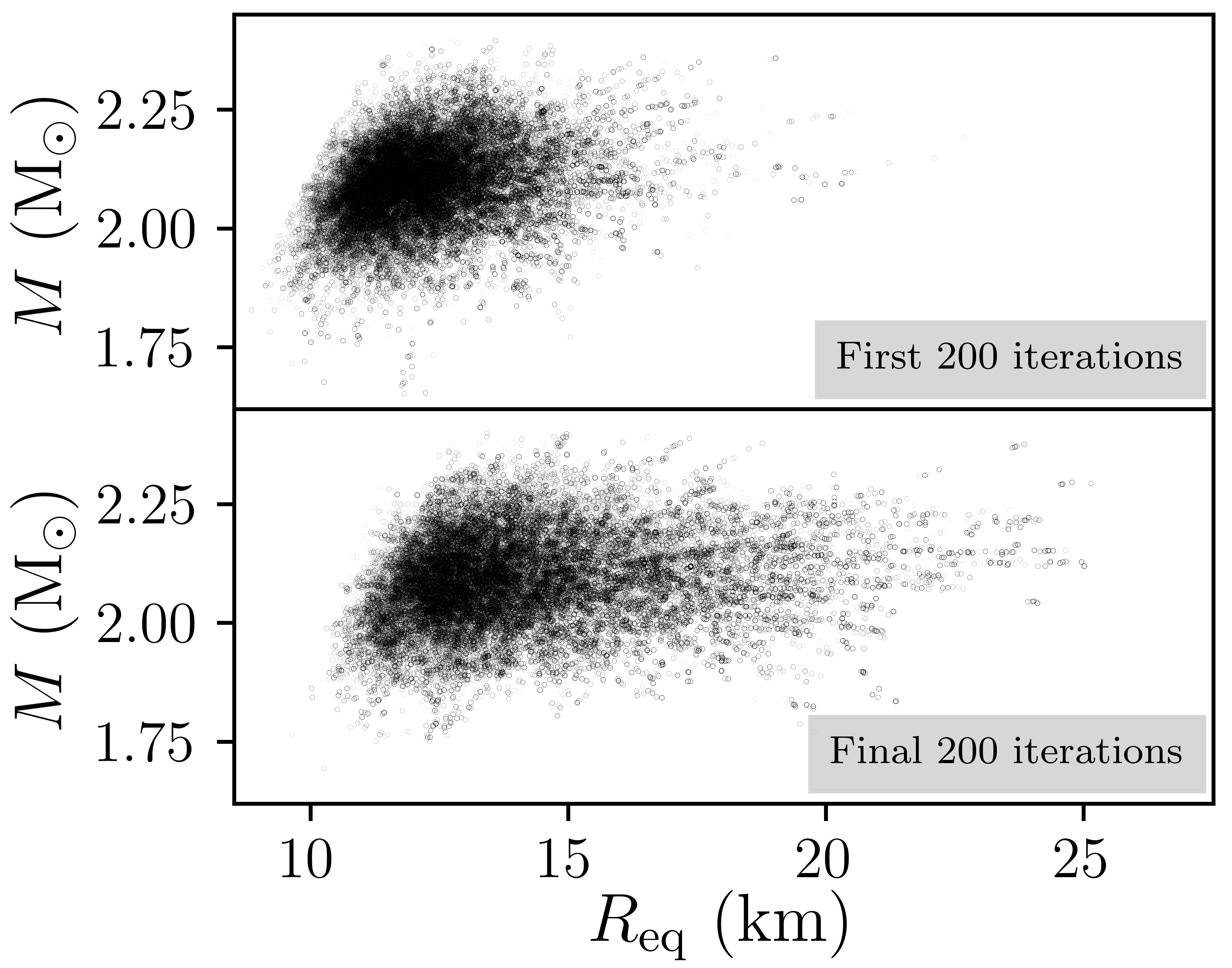}
  \end{minipage}\hfill
  \begin{minipage}[c]{0.52\textwidth}
\vspace{-0.02truein}
\caption{Another view of the evolution of the sampling in our analysis of the \textit{NICER} and \textit{XMM-Newton} data on PSR~J0740$+$6620.  The top panel shows the positions of the walkers in mass-radius space during the first 200 iterations of our PT-emcee sampling; these points were drawn from the posterior of an Nlive=1000, efficiency=0.01 MultiNest run, using a kernel density estimate.  The bottom panel shows the positions of the walkers during the final 200 iterations, after the PT-emcee sampling had converged.  These panels correspond to the beginning and end of the left panel of Figure~\ref{fig:convergence}, and demonstrate the difference in the sampling between the two stages.}       
\label{fig:conv2}
  \end{minipage}
\end{figure*}

\subsection{Convergence of Sampling}
\label{sec:convergence}

As we discussed in Section~\ref{sec:parameter-estimation}, we started our PT-emcee sampling of the posterior using an initial distribution of points that had previously been obtained by using MultiNest (with Nlive=1000 and efficiency=0.01) to sample the posterior; as usual, MultiNest had run until the evidence estimate converged.  The left panel of Figure~\ref{fig:convergence} shows the evolution of the $-1\sigma$, median, and $+1\sigma$ values of the radius posterior as our PT-emcee sampling of the \textit{NICER}+\textit{XMM-Newton} data using the nominal calibration progressed, starting with the initial parameter distribution provided by MultiNest.  Each point in this plot indicates the estimates of these three radii given by a single set of 20,000 samples; the individual sets are contiguous and non-overlapping. The evolution shown was computed using approximately 1700 such sample sets; we show another view of the progressive mass-radius sampling in Figure~\ref{fig:conv2}.  The left panel in Figure~\ref{fig:convergence} shows that as the sampling by PT-emcee proceeded, the values of all three characteristic radii increased rapidly at first, indicating that the sampling achieved by MultiNest was insufficient and that the estimated values of the parameters it provided were inaccurate. Eventually the values of the three parameters ceased increasing and begin wandering up and down stochastically by small amounts, indicating that the PT-emcee estimates of the three radii had converged. 

The dashed vertical line indicates the beginning of the sample sets that we used to construct our final posterior distributions (a total of $10^7$ samples).  The effective number of independent samples can be estimated by dividing the total number of samples by the average integrated autocorrelation time.  Using typical methods for calculating the autocorrelation time for each walker (e.g., \citealt{sokal1997}) yields an estimate of $\sim 4$ iterations, suggesting a total of $\sim 2.5\times 10^6$ independent samples.  From sample set to sample set in the converged part of the evolution, the median, $\pm 1\sigma$, and $\pm 2\sigma$ radii vary with a standard deviation of only a few hundredths of a kilometer.

The right panel of Figure~\ref{fig:convergence} shows the evolution of the three radii in our PT-emcee sampling of the model for the \textit{NICER}+\textit{XMM-Newton} data when a calibration parameter is included.  Again we show the three radius values computed using contiguous and non-overlapping sets of 20,000 samples.  This sampling was initiated using kernel density estimation based on 10\% of the final posterior samples from the PT-emcee analysis of \textit{NICER}+\textit{XMM-Newton} data performed using the nominal \textit{XMM-Newton} calibration.  As a result, the initial values of the three radii did not match perfectly the final values of the three radii in the left panel, indicating that the initial posterior distribution in the right panel does not quite match the final posterior distribution from the run shown in the left panel.  However, as Figure~\ref{fig:convergence} shows, these three radius estimates rapidly approached the values they had at the end of the run shown in the left panel, indicating that as the PT-emcee sampling progressed, the posterior distribution converged quickly to the final posterior distribution in the run shown in the left panel.  

We note that both our MultiNest and our PT-emcee analyses of only the \textit{NICER} data, and of the \textit{NICER} plus \textit{XMM-Newton} data that were performed using the nominal \textit{XMM-Newton} calibration, assumed a mass prior of $2.11\pm 0.09~M_\odot$, consistent with a preliminary radio-based mass measurement.  The posterior function that was sampled to produce the three radius values plotted in the left panel of Figure~\ref{fig:convergence} was computed assuming this prior.  However, the posterior function that was sampled to produce the three radius values plotted in the right panel assumed a different mass prior of $2.08\pm 0.09~M_\odot$, consistent with the current radio-based mass measurement. As a result, the radius values plotted in the right panel of the figure differ by about 1.5\% from the radius values plotted in the left panel.  The numbers in the tables and figures that summarize the posteriors produced by our analyses of the \textit{NICER}-only and the \textit{NICER} and \textit{XMM-Newton} data assuming the nominal calibration have been re-weighted using the updated mass.

\subsection{Effect of Priors}
\label{sec:priors}

\begin{deluxetable*}{c|c|c|c}
    \tablecaption{Summary of Inferred Radii Using Different Priors}
\tablewidth{0pt}
\tablehead{
      \colhead{Prior} & \colhead{$-1\sigma$~(km)} & \colhead{Median (km)} & \colhead{$+1\sigma$~(km)} 
    \label{tab:priors}
 }
\startdata
      \hline
      \textbf{\textit{NICER}+\textit{XMM-Newton} (featured result)} & \textbf{12.209} & \textbf{13.713} & \textbf{16.326} \\
      \hline
      \textit{NICER}+\textit{XMM-Newton}, $A_{\rm XMM}$ & 12.153 & 13.705 & 16.298 \\
      \hline
       \textit{NICER}+\textit{XMM-Newton}, $\theta_{\rm obs}$=radio & 12.236 & 13.750 & 16.416 \\
      \hline
       \textit{NICER} only & 10.382 & 11.512 & 13.380 \\
      \hline
\enddata
\tablecomments{Comparison of $\pm 1\sigma$ radii inferred using different priors on the \textit{NICER} plus \textit{XMM-Newton} data sets, and for the \textit{NICER}-only analysis.  See text for details and note that the top, boldfaced numbers are from the analysis that we use in Section~\ref{sec:EOS} for our EOS inference.  When the \textit{NICER} and \textit{XMM-Newton} data sets are analyzed jointly the $-1\sigma$ radius, which is important for EOS analysis (see Section~\ref{sec:EOS}), is robust against different priors.}
\end{deluxetable*}

As summarized in Table~\ref{tab:priors}, our posteriors are robust against several different choices of priors.  In Table~\ref{tab:priors} we compare the $\pm 1\sigma$ radii when we analyze the \textit{NICER} and \textit{XMM-Newton} data using our standard priors with the $\pm 1\sigma$ radii when we (1)~include a parameter $A_{\rm XMM}$ that models the relative calibration of \textit{NICER} and \textit{XMM-Newton} (as discussed before, our flat prior $0.9\leq A_{\rm XMM}\leq 1.1$ is broader that the expected uncertainty around $A_{\rm XMM}=1$ by a factor $\sim 2$), and (2)~restrict our inclination prior to being flat within $\pm 0.5^\circ$ of the radio measurement, rather than our standard $\pm 5^\circ$.  In each case the $-1\sigma$ radius (which has especially important implications for the EOS; see Section~\ref{sec:EOS}) is close to our standard value.  For comparison we also show in this table the $\pm 1\sigma$ radii when we use only the \textit{NICER} data, with no assumed knowledge of the \textit{NICER} background.
\bigskip

\subsection{Adequacy of the Models}
\label{sec:adequacy}

\begin{figure*}[ht!]
  \begin{minipage}[c]{0.5\textwidth}
    \includegraphics[width=\textwidth]{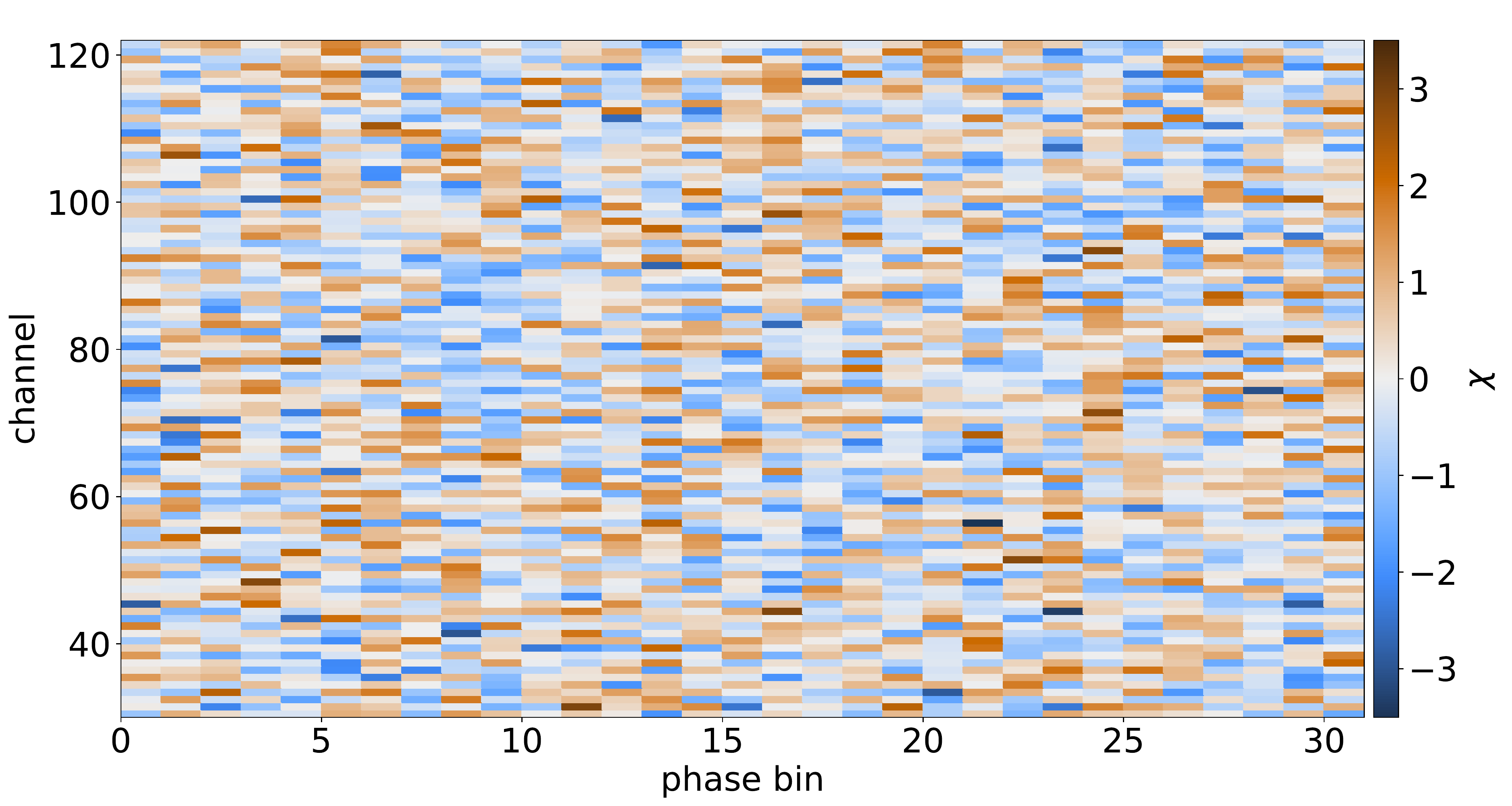}
  \end{minipage}\hfill
  \begin{minipage}[c]{0.45\textwidth}
    \caption{The value of $\chi$ in each of the 3008 \textit{NICER} phase-energy bins (32~phase bins and 94~energy channels), for the best fit of our energy-resolved waveform model to the \textit{NICER} plus the \textit{XMM-Newton} data.  As is expected for a good fit, no patterns in the values of $\chi$ are evident as a function of phase or energy.  The $\chi^2/{\rm dof}$ is 2912.37/2901, which has a probability of 0.437 if the model is correct.}       
\label{fig:chi-plot}
  \end{minipage}
\end{figure*}

\begin{figure*}[ht!]
\begin{center}
\vspace*{-1.5truein}
  \resizebox{0.9\textwidth}{!}{\includegraphics{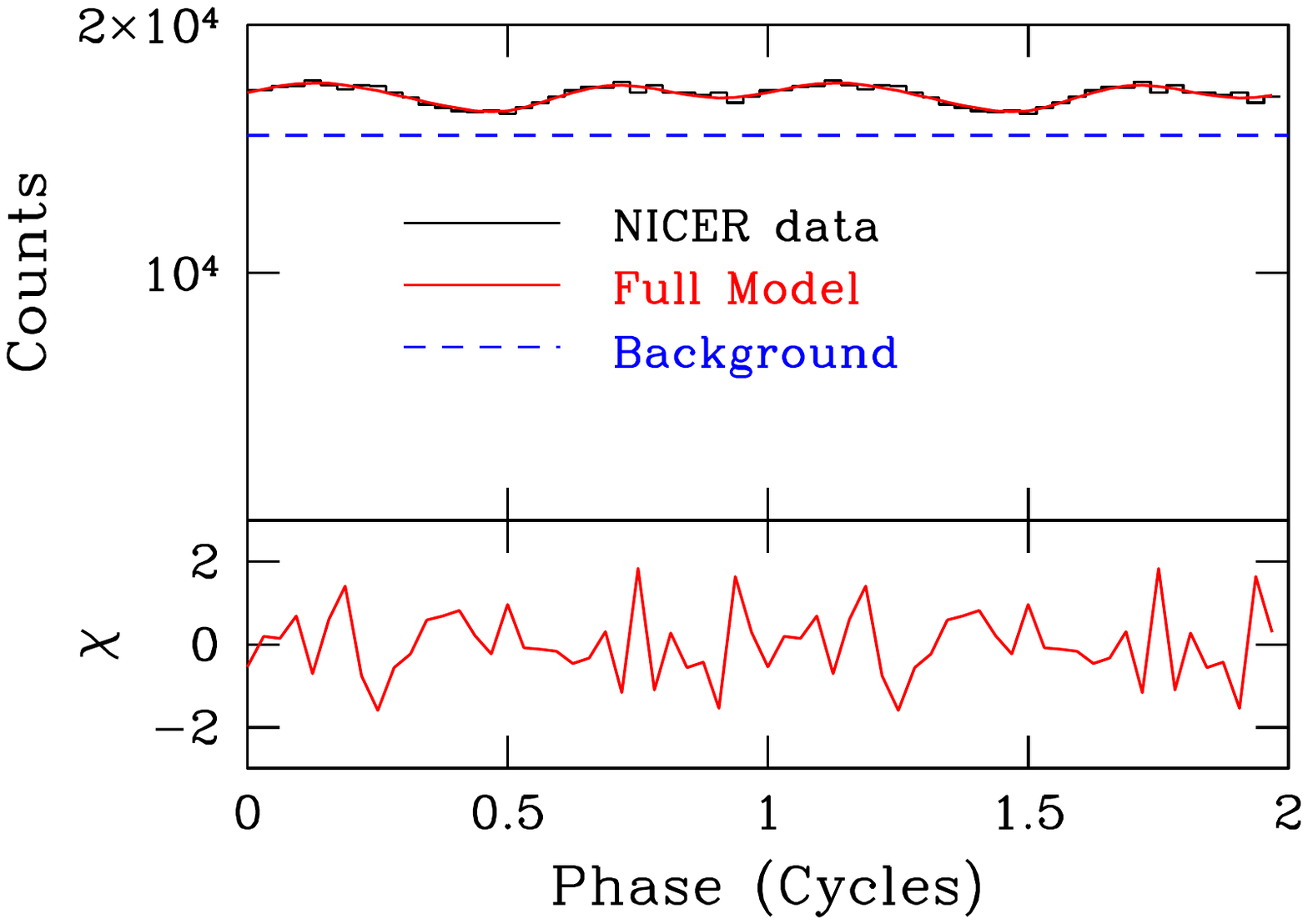}}
\vspace{-3.5truein}
   \caption{
Top: Comparison of the 32-phase bolometric waveform constructed using the \textit{NICER} data on PSR~J0740$+$6620 with the bolometric waveform model that best fits the \textit{NICER} plus the \textit{XMM-Newton} data.  The dashed blue line shows the unmodulated background that was added to the counts produced by the two hot spots as part of the fitting procedure (see Section~\ref{sec:background}).  Bottom: The resulting value of $\chi$ as a function of phase. The $\chi^2/{\rm dof}$ is 21.85/18, which has a probability of 0.239 if the model is correct.
    }
\label{fig:bolo-waveform+residuals}
\end{center}
\end{figure*}

\begin{figure*}[ht!]
  \begin{minipage}[c]{0.5\textwidth}
    \includegraphics[width=\textwidth]{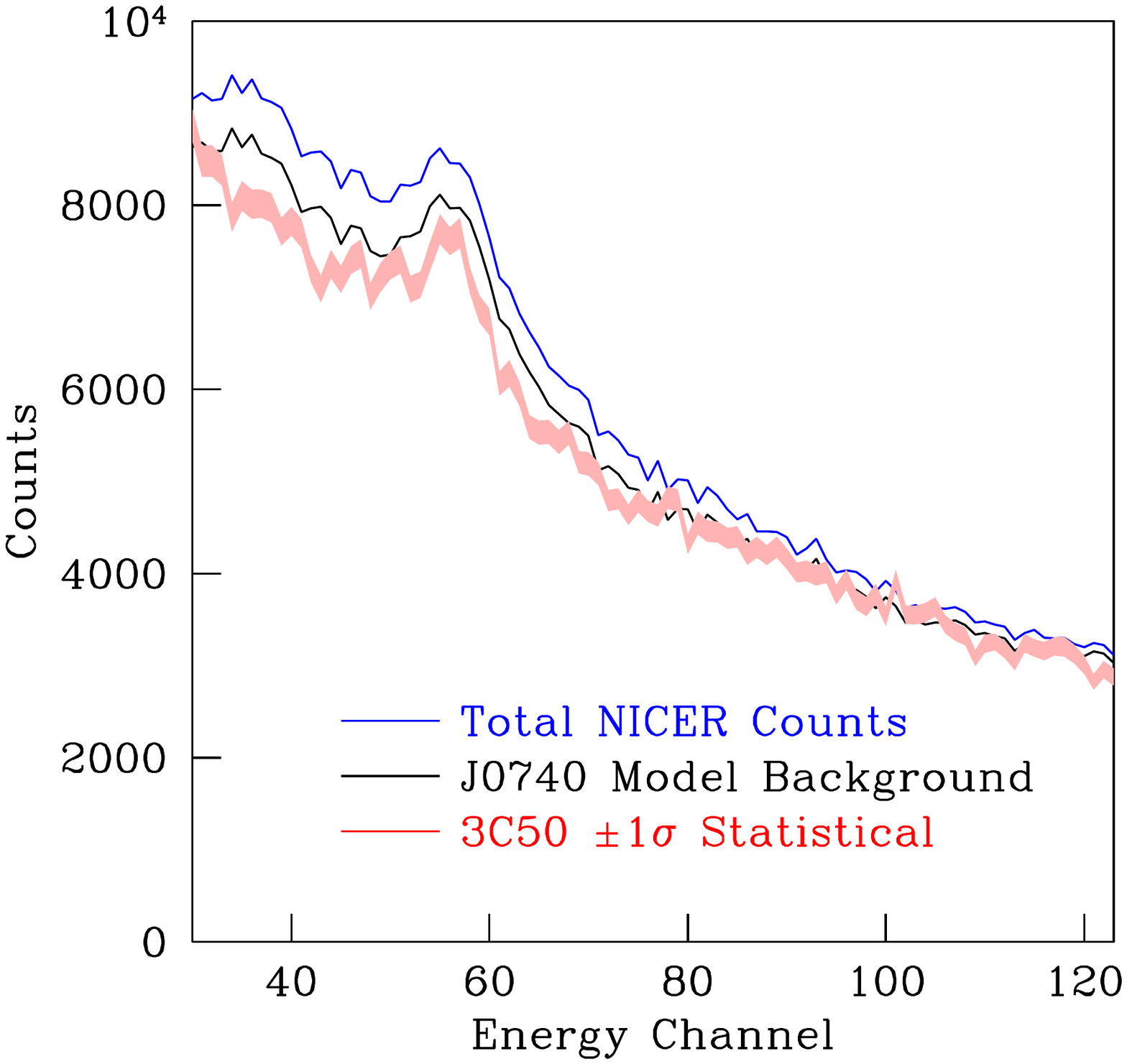}
  \end{minipage}\hfill
  \begin{minipage}[c]{0.45\textwidth}
\vspace{-0.3truein}
    \caption{The inferred non-spot counts as a function of \textit{NICER} energy channel in our best model (black line) compared with the $\pm 1\sigma$ range of the 3C50 background model (pink band) of Remillard et al. (2021, submitted).  For comparison, we also show the total \textit{NICER} counts per channel (blue line).  Above channel 80 our inferred background is consistent with the 3C50 model.  The excess in our inferred background at lower energies is likely due to the presence of other sources in the \textit{NICER} field, which can be seen in \textit{XMM-Newton} images.}       
\label{fig:backgrounds}
  \end{minipage}
\vspace*{-1.0truein}
\end{figure*}

As in \citet{2019ApJ...887L..24M}, we performed $\chi^2$ tests to assess the adequacy of our models in describing the data.  This is a one-way test: if $\chi^2\sim{\rm dof}$ for ${\rm dof}$ degrees of freedom this does not guarantee that the model is correct, but if $\chi^2$ is enough larger than ${\rm dof}$ that the probability is very low, it indicates that the model is deficient.  We focus on the results of the joint fit to the \textit{NICER} and \textit{XMM-Newton} data using the nominal \textit{XMM-Newton} calibration, given that the \textit{XMM-Newton} data refine the solution substantially and that allowance for a shift in the effective area of \textit{XMM-Newton} does not make a significant difference to the posterior credible regions.  

In Figure~\ref{fig:chi-plot} we show the value of $\chi$ over each of the \textit{NICER} phase-channel bins we used in our analysis, for our best fit to the \textit{NICER} plus the \textit{XMM-Newton} data.  There are no obvious patterns, and the $\chi^2/{\rm dof}$ is 2912.37/2901 for the \textit{NICER} data alone, which has a probability of 0.437 if the model is correct.  In Figure~\ref{fig:bolo-waveform+residuals} we compare to the data our bolometric model waveform for our best fit, and also display the residuals.  Here $\chi^2/{\rm dof}=21.85/18$ (probability of 0.239) for the \textit{NICER} data alone.  

Our final comparison to the \textit{NICER} data is in Figure~\ref{fig:backgrounds}, which compares the non-spot background in our best model (black line) with the blank-sky \textit{NICER} background inferred using the 3C50 model of Remillard et al. (2021, submitted; see https://heasarc.gsfc.nasa.gov/docs/nicer/tools/nicer\_bkg\_est\_tools.html).  Because the 3C50 model is for a blank sky, i.e., directions where there are no known sources, it should provide a lower limit to the background in the direction of a particular source such as PSR~J0740$+$6620.  Indeed, we see that our estimated background counts, while consistent with the 3C50 model at energy channels $\approx 80$ and higher, is above the model at lower energy channels.  This is likely to be due to particular sources within the \textit{NICER} field of view that can be identified in the \textit{XMM-Newton} image of the field (see \citealt{2021..............W}).  

Our model is therefore a good fit to the \textit{NICER} data.

\begin{figure*}[ht!]
\begin{center}
\vspace*{-2.0truein}
  \resizebox{1.0\textwidth}{!}{\includegraphics{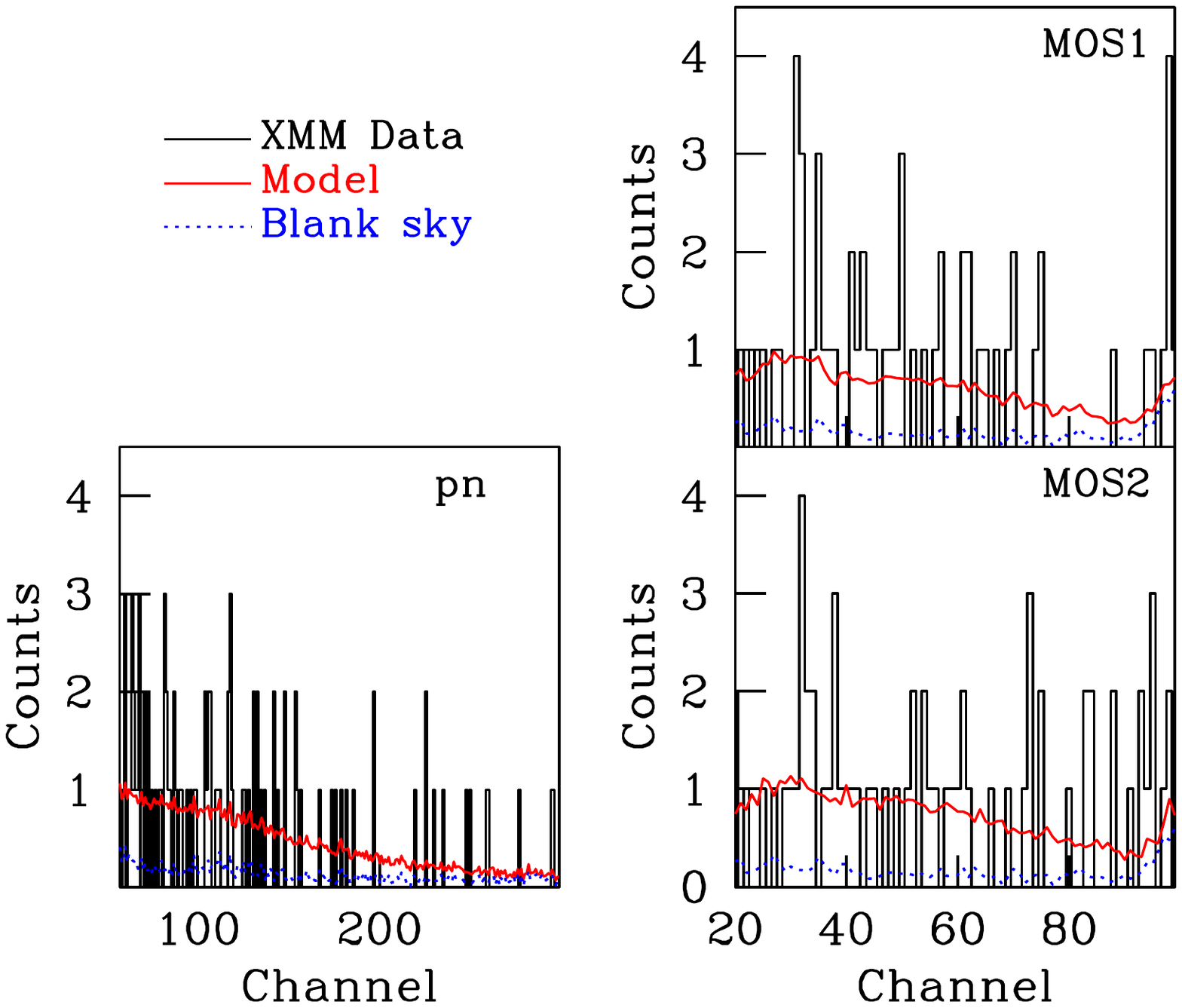}}
\vspace{-3.0truein}
   \caption{
Comparison of the data from the three \textit{XMM-Newton} cameras (labeled at the top of each plot) with our best \textit{NICER}+\textit{XMM-Newton} model.  In each plot the black histogram is the data, the dotted blue line is the blank-sky background scaled to the duration of the observations, and the red line is the best model including the background.  In each case the energy channel range is the range we use in our analysis.  As described in the text, our model provides a good fit to the data.
    }
\label{fig:XMMchannel}
\end{center}
\end{figure*}

In Figure~\ref{fig:XMMchannel} we compare our best \textit{NICER}+\textit{XMM-Newton} model with the \textit{XMM-Newton} data.  Because there are few counts per energy channel, we evaluated the quality of our fit by generating $10^5$ synthetic data sets by performing Poisson draws using our best model (including the blank-sky counts) in every energy channel in all three instruments.  We then compared our model to the J0740 data, and placed that in the context of the comparison of our model to the synthetic data sets, in two ways:

\begin{itemize}

\item We computed the total log likelihood of all three \textit{XMM-Newton} data sets for PSR~J0740$+$6620 given our model.  We also computed the distribution of the total log likelihoods of each synthetic data set given our model.  The total log likelihood of the real data is at the 75th percentile of the log likelihoods from the synthetic set (98th percentile for the pn data alone; 17th percentile for the MOS1 data alone; and 12th percentile for the MOS2 data alone).

\item We also calculated the Kolmogorov-Smirnov (K-S) probability that the cumulative counts in our model, as a function of the energy channel, are drawn from the same distribution as the PSR~J0740$+$6620 data.  We then calculated the distribution of the K-S probabilities for the synthetic data set.  We found that K-S probability for the \textit{XMM-Newton} pn data on PSR~J0740$+$6620 is at the 10th percentile of the corresponding probabilities for the synthetic data sets; the K-S probability for the MOS1 data on PSR~J0740$+$6620 is at the 50th percentile; and the K-S probability for the MOS2 data on PSR~J0740$+$6620 is at the 31st percentile.

\end{itemize}

As a final check, we note that the \textit{XMM-Newton} DDT data have 94 counts in the pn camera in the energy range we investigate, 59 counts in the MOS1 camera, and 67 counts in the MOS2 camera.  In our best fit, we have respectively 114.88 counts (1.95 standard deviations high), 49.85 counts (1.30 standard deviations low), and 59.78 counts (0.93 standard deviations low).  Neither the total count numbers nor the comparisons with synthetic data indicate a problem.  Thus, at least based on these tests, our model is fully consistent with all three \textit{XMM-Newton} data sets.

\subsection{Potentially important differences between our analysis and that of \citet{2021..............R}}
\label{sec:differences}

The parallel analysis of the \textit{NICER} and \textit{XMM-Newton} data by \citet{2021..............R} finds a smaller credible interval for the radius estimate and, most important for the inferred constraints on the EOS, a value for the radius at the $-1\sigma$ contour that is $\approx 0.8$~km smaller than the value we report here.  There are a number of differences between their analysis procedure and ours.  For example, \citet{2021..............R} assume that the inclination of the observer's line of sight to the pulsar's rotational axis is identical to the measured inclination of the observer's line of sight to the orbital axis of the binary system, whereas we allow the inclination of the observer's line of sight to the pulsar's rotational axis to differ by as much as $\pm 5^\circ$ from the measured inclination of the observer's line of sight to the orbital axis. However, as we have discussed in Section 4.4, this difference between the priors on the inclination of the observer's line of sight to the pulsar's rotational axis used in the two analyses does not produce a significant difference in the value of the radius at the $-1\sigma$ credibility contour.  Another unimportant difference is the ionization fraction of our hydrogen atmospheres: we use tables that allow for the possibility of partial ionization, whereas \citet{2021..............R} used atmospheres with fully ionized hydrogen.  As we indicate in Section~\ref{sec:modeling-the-emission}, we found negligible changes in the radius posteriors when we used fully ionized hydrogen atmospheres.

There appear to be four differences in the analysis procedures used by the two groups that could contribute to the difference of $\approx 0.8$~km in the values of the radius at the $-1\sigma$ credibility contour that the two groups report in their headline results ($R = 12.2$~km at $-1\sigma$ reported here and $R = 11.4$~km at $-1\sigma$ reported by \citealt{2021..............R}):

\begin{enumerate}

\item We allow the relative calibration between \textit{NICER} and \textit{XMM-Newton} to deviate by up to $\pm 10$\% from its nominal value, a deviation that is a factor $\sim 2$ times larger than the measured deviation (again see slide 7 of http://iachec.org/wp-content/presentations/2020/NICER-CrossCal-IACHEC-Markwardt-2020b.pdf and focus in particular on the energy range $0.3-1$~keV that is relevant for the present analysis; here the deviation is $<4$\%).  In contrast, \citet{2021..............R} allow the relative calibration to deviate by several tens of percent, which is many times larger than the measured deviation.  Elsewhere in their paper, \citet{2021..............R} show that when they restrict the relative calibration to a range consistent with the measured range, they find $R = 11.75$~km at the $-1\sigma$ contour. Thus, when they require the relative calibration between \textit{NICER} and \textit{XMM-Newton} to have a value consistent with the measured values, the difference between their value of the radius estimate at the $-1\sigma$ contour and our value is reduced by almost a factor of two.

\item When estimating the radius of PSR~J0740$+$6620, we allowed $R_{\rm eq}c^2/(GM)$ to be as large as 8.0 (corresponding to $R_{\rm eq}\approx 24.5$~km at $M=2.08~M_\odot$) and found that the posterior radius probability density is negligible near this boundary. Hence, this choice for the boundary of our search volume did not affect our estimate of the radius of PSR~J0740$+$6620.  In contrast, when \citet{2021..............R} estimated the radius of PSR~J0740$+$6620, they sampled the radius parameter space only at radius values $<16$~km.  To investigate the results we would have found if we had sampled the radius parameter space only at $R<16$~km as they did, we required $R<16$~km in a test analysis and found $R = 12.06$~km, $13.28$~km, and $14.82$~km at the $-1\sigma$ contour, the median value, and the $+1\sigma$ contour.  Even when we imposed $R<16$~km and also allowed the relative calibration factor $A_{\rm XMM}$ to vary freely in the range $0.9 < A_{\rm XMM} < 1.1$, a factor $>2$ deviation larger than its measured value, we found that this had only a very small effect on the position and width of the credible interval of the radius, which changed only slightly to $R = 12.00$~km, $13.28$~km, and $14.80$~km at the $-1\sigma$ contour, the median value, and the $+1\sigma$ contour.  Thus, these two differences between our analysis procedure and the analysis procedure used by \citet{2021..............R} (the different breadths of the allowed deviation of the relative calibration of the \textit{NICER} and \textit{XMM-Newton} instruments and artificially limiting or not limiting the sampling of the radius parameter volume) account for $\approx 0.55$~km of the 0.8~km difference between the value of the radius at the $-1\sigma$ contour reported by \citet{2021..............R} and the value of the radius at the $-1\sigma$ contour that we found. 

\item We used PT-emcee to thoroughly sample the multi-dimensional posterior, using MultiNest only to create an initial rough sample that we then used to speed up the sampling using PT-emcee.  In our analysis of synthetic pulse waveform data constructed to mimic the actual \textit{NICER} and \textit{XMM-Newton} data (see Section~\ref{sec:synthetic}), we found that when we used only MultiNest, even with values of the MultiNest parameters chosen to try to achieve thorough sampling, the credible regions it generated were too small, i.e., they too often failed to include the values of the pulse waveform parameters that had been used to generate the synthetic pulse waveform data.  In contrast, when we used PT-emcee we found credible regions that were consistent with the values of the parameters chosen when generating the synthetic data (see Table~\ref{tab:synth} for more details).  Figure~\ref{fig:convergence} shows the detailed evolution of the $-1\sigma$, median, and $+1\sigma$ radii throughout the course of our PT-emcee sampling.  \citet{2021..............R} used only MultiNest for all of their analyses and found that their posteriors continued to broaden with more thorough sampling for all of the live point numbers that they explored.  

\item We treated the distribution of the blank-sky counts in each energy channel as a realization of a Poisson distribution for the observed number of counts. \citet{2021..............R} instead assumed a flat prior for the distribution of the blank-sky counts in each energy channel, centered on the flux implied by the observed number of counts with a variety of widths equal to various multiples of the standard deviation of the observed number of counts.

\end{enumerate}

\section{IMPLICATIONS FOR THE EOS}
\label{sec:EOS}

Our aim is to combine multiple lines of data relevant to the properties of cold, catalyzed, dense matter, using a Bayesian procedure that is both statistically rigorous and computationally practical.  In this section we divide our approach into two subtasks: (1)~likelihood computation and parameter estimation for a given EOS family (Section~\ref{sec:statistical}), and (2)~different approaches to generating candidate EOS (Section~\ref{sec:models}).  Finally, we show in Section~\ref{sec:results} that with the inclusion of the mass and radius estimates for PSR~J0030$+$0451 and PSR~J0740$+$6620, the constraints on the EOS depend relatively weakly on the EOS parameterization assumed in the analysis, up to a few times nuclear saturation density. 

\subsection{Statistical Method}
\label{sec:statistical}

There is a growing number of papers on EOS inference from astronomical and/or laboratory data (e.g., \citealt{2015PhRvD..91d3002L,2015PhRvD..92b3012A,2016EPJA...52...69A,2017ApJ...850L..19M,PhysRevLett.120.172703,PhysRevLett.120.261103,2018ApJ...852L..25R,2018ApJ...852L..29R,2018MNRAS.478.1093R,PhysRevC.98.045804,2019MNRAS.485.5363G,2019ApJ...876L..31K,2019PhRvD..99h4049L,PhysRevD.100.023015,2020NatAs...4..625C,2020EPJST.229.3663C,2020Sci...370.1450D,2020PhRvC.102d5807L,2020PhRvC.102e5803E,2020PhRvD.101l3007L,PhysRevResearch.2.033514,2020ApJ...893L..21R,2020ApJ...893...61Z,2021PhRvC.103c5802X}).  Here we follow the procedure described in \citet{2020ApJ...888...12M}.  Their procedure is fast and flexible enough to incorporate multiple types of data.

To elaborate, suppose that we have $i$ different types of measurements, which might include laboratory experiments, or the measurement of a high mass from a pulsar, or the measurement of a tidal deformability, a radius, a moment of inertia, and so on.  Suppose also that for measurement type $i$ we have $j=1, 2,\ldots,j(i)$ independent measurements of that type; these could be different measurements of the same source, or measurements of different sources\footnote{For example, here we present our radius measurements, which are based on X-ray data, using a prior on the mass given by radio observations.  This prior, however, is obtained by starting with a flat mass prior and then updating the mass distribution using the radio data.  Thus our procedure is equivalent to starting with a flat mass prior and then computing the joint likelihood of the radio and X-ray data given candidate masses and radii.}.  We wish to evaluate a set $k$ of EOS; in the next section we discuss the selection of $k$.  If the likelihood of data set $(i,j)$ given EOS $k$ is ${\cal L}_k(i,j)$, then the likelihood of all of the data sets given EOS $k$ is simply the product of the likelihoods:
\begin{equation}
{\cal L}_k=\prod_i\left[\prod_{j=1}^{j(i)}{\cal L}_k(i,j)\right]\; .
\end{equation}
Note that in some cases one may need to marginalize over extra parameters.  For example, to compute the likelihood of a \textit{NICER} posterior in mass and radius it is necessary to integrate over the prior for the central density, which is not the same for all EOS because different EOS have different maximum stable central densities. 

Given the calculation of the likelihood ${\cal L}_k$ of the full set of data assuming a specific EOS $k$ (see \citealt{2020ApJ...888...12M} for more details), then as usual in a Bayesian analysis the posterior probability $P_k$ of the EOS is proportional to the prior probability $q_k$ of the EOS times the likelihood:
\begin{equation}
P_k\propto q_k{\cal L}_k\; .
\end{equation}
We could use this procedure to compare individual EOS from the literature.  However, our primary goal is to produce a posterior probability distribution for, say, the pressure $P$ as a function of the total energy density $\epsilon$ or of the number density $n$ of baryons.  To do this, we simply calculate the pressure at specified $\epsilon$ or $n$ for each EOS $k$, weight it by $P_k$, and then sum over all $k$ to find the final distribution of $P$ at a given $\epsilon$ or $n$.

For each specific EOS selected from a given EOS family we need to compute the likelihood of all relevant data sets for that EOS.  We largely follow the procedure of \citet{2019ApJ...887L..24M} in our treatment of our data sets.  In particular, we apply kernel density estimation (other methods exist, e.g., random forest regressors; see \citealt{PhysRevD.100.103009}) to produce an estimated continuous probability distribution from our discrete samples.  When we do so, we use the standard bandwidth from \citet{silverman1986} for our samples in mass and tidal deformability, but multiply this bandwidth by 0.1 for our mass-radius samples because we found previously that the standard bandwidth significantly broadened the radius posterior (see \citealt{2019ApJ...887L..24M} for more details, and see \citealt{2020PhRvD.101f3007E} for a more detailed approach to selecting the bandwidth).  We modify the procedure of \citet{2020ApJ...888...12M} in two ways:

\begin{enumerate}

\item In \citet{2020ApJ...888...12M} and \citet{2019ApJ...887L..24M} the prior on the central density was flat between the density that produces a $1~M_\odot$ star and the density that produces the maximum-mass nonrotating neutron star for the EOS under consideration.  Because the central density changes rapidly near the maximum mass, this gives relatively greater prior weight to higher-mass stars.  Here our prior on the central density is instead quadratic in the central density between the $1~M_\odot$ density, $\rho_{\rm min}$ and the maximum-mass density $\rho_{\rm max}$.  That is, we select uniformly in $0\leq x\leq 1$, where the central density is $\rho_c=\rho_{\rm min}+x^2(\rho_{\rm max}-\rho_{\rm min})$.  This produces a more even prior distribution on the masses.  However, we assign zero prior weight to any densities between $\rho_{\rm min}$ and $\rho_{\rm max}$ that produce unstable stars; in this way we explicitly treat EOS that have two stable ranges of central density separated by an unstable range of central density (these are sometimes called ``twin stars"), when they arise.  In practice, the different prior distribution for the central density does not change the EOS posteriors significantly compared with the approach in \citet{2019ApJ...887L..24M}.

\item Because we include the high-mass binary merger GW190425 \citep{2020ApJ...892L...3A} in our analysis, and the total mass of the binary is high enough that one of the compact objects might have been a black hole, we treat tidal deformabilities differently than in \citet{2020ApJ...888...12M} and \citet{2019ApJ...887L..24M} (although we also note that the large distance to and high mass of GW190425 mean that it adds only a small amount of information).  Our approach is to select the central density of the lower-mass star, up to the central density such that the implied mass of the other star equals that of the lower-mass star.  Given the mass of the lower-mass star, and given that the chirp mass (equal to $(m_1m_2)^{3/5}/(m_1+m_2)^{1/5}$ for stellar masses $m_1$ and $m_2$ in a binary) is known with high precision, we know the mass of the higher-mass object to the precision that we know the chirp mass.  We marginalize over the chirp mass as in Equation~(15) of \citet{2020ApJ...888...12M}.  If for the EOS and central density under consideration the higher-mass object is a neutron star, then we compute the tidal deformabilities of both stars, using the same EOS, following the prescription of \citet{2008ApJ...677.1216H} (see the erratum at \citealt{2009ApJ...697..964H}).  If instead for the EOS and central density under consideration the higher-mass object is a black hole, then we assume that its tidal deformability is zero \citep{2009PhRvD..80h4018B,2009PhRvD..80h4035D}.  Note that for rotating black holes the tidal deformability may not be exactly zero (see \citealt{2020arXiv200700214L,2020arXiv201007300C,2020arXiv201015795L} for an ongoing discussion), but it is small enough that the difference from zero is not important for our calculations.

\end{enumerate}

To implement our analysis, we need a mechanism to select a set of EOS.  We now discuss our three different approaches to this selection.

\subsection{EOS Models}
\label{sec:models}

There exist many different nuclear physics computations of the EOS of cold, catalyzed matter both below and above nuclear saturation density.  It is usually agreed that our understanding of matter below saturation density is fairly good, based on the idea that we can perform laboratory experiments that probe this density regime.  One caveat is in order: the matter that can be examined in laboratories has roughly equal numbers of neutrons and protons (the most asymmetric stable nuclei, such as $^{208}$Pb, have roughly 50\% more neutrons than protons).  In contrast, neutron star matter in the vicinity of saturation density is more than 90\% neutrons.  Thus although from the standpoint of density the regime is familiar, theoretical arguments not directly supported by experiment are necessary to extrapolate to very asymmetric matter.  

Nonetheless, we will follow standard procedure and assume that we know the EOS up to a threshold density, which we take to be half of saturation density because this is approximately the crust-core transition density \citep{2013ApJ...773...11H}.  Note that some calculations suggest that different treatments of the crustal EOS could introduce radius uncertainties as large as 0.3~km \citep{2016PhRvC..94c5804F,2020CQGra..37b5008G}.  Below half of nuclear saturation density we use the QHC19 EOS \citep{2019ApJ...885...42B}.  Above half of nuclear saturation density, we then need to model the EOS, where there are well motivated and carefully constructed models, but where the models have not been validated by laboratory measurements.

A necessary output of our EOS analysis is an understanding of how strongly our conclusions depend on the type of EOS model that we employ.  Therefore, in the rest of this section we discuss the three EOS model families that we use in our analysis.  First we present two parameterizations which have been widely used in the literature.  We then turn our attention to a newer method using Gaussian processes (GPs) that was introduced in this context by \citet{2019PhRvD..99h4049L}.  

\subsubsection{Parametric models}
\label{sec:parametric}

The most common approach to EOS inference has used parameterized models.  That is, we assume that there is some set of parameters ${\bf\alpha}$ that completely describe the EOS at high densities.  Given priors on ${\bf\alpha}$, we can generate a large number of sample EOS for which we then compute the likelihood of the assembled data.  As discussed above, this allows us to compute posteriors for $P(\epsilon)$ or $P(n)$.  The two parameterizations we use are:

\textit{Piecewise polytrope.}---In this model, above a number density $n=n_s/2$ the pressure is given by a series of polytropes: $P=K_1n^{\Gamma_1}$ for $n_s/2\leq n\leq n_2$, $P=K_2n^{\Gamma_2}$ for $n_2\leq n\leq n_3$, and so on, where the coefficients $K_i$ are chosen to ensure continuity of pressure between density segments.  Variants of this model differ in the number of segments, the priors on the polytropic indices $\Gamma_i$, and whether the transition densities $n_1,n_2,\ldots$ are fixed or can vary.  We use the following assumptions (compare with \citealt{2019ApJ...887L..24M}, which used the same assumptions except for restricting the first polytropic index to the range $[2,3]$): the polytropic index from $n_s/2$ to $n_2\in[3/4,5/4]n_s$ is $\Gamma_1\in[0,5]$; from $n_2$ to $n_3\in[3/2,5/2]n_s$ is $\Gamma_2\in[0,5]$; from $n_3$ to $n_4\in[3,5]n_s$ is $\Gamma_3\in[0,5]$; from $n_4$ to $n_5\in[6,10]n_s$ is $\Gamma_4\in[0,5]$; and from $n_5$ to $\infty$ is $\Gamma_4\in[0,5]$ (all priors are uniform in the listed range).  When $\Gamma$ is close to zero, the pressure is nearly independent of the density, which is what one expects near a phase transition and which can produce the twin stars discussed earlier.  If at some density below $n_5$ the implied adiabatic sound speed $c_s=\left(dP/d\epsilon\right)^{1/2}>c$, the speed of light, then we set $dP/d\epsilon=c^2$ at that density.

\textit{Spectral parameterization.}---A different parameterization was introduced by \citet{2010PhRvD..82j3011L} and \citet{2018PhRvD..97l3019L}.  In this approach, the polytropic index is a continuous function of the pressure, $\Gamma(P)$:
\begin{equation}
\Gamma(P)=\exp\left(\sum_i\gamma_ix^i\right)\; .
\end{equation}
Here $x\equiv\log(P/P_0)$ and $P_0$ is the pressure at $n_s/2$.  We follow \citet{2018PhRvD..98f3004C}, \citet{2018PhRvL.121p1101A}, and \citet{2019ApJ...887L..24M} in expanding to $O(x^3)$, and in using uniform priors $\gamma_0\in[0.2,2]$, $\gamma_1\in[-1.6,1.7]$, $\gamma_2\in[-0.6,0.6]$, and $\gamma_3\in[-0.02,0.02]$.  As with the piecewise polytrope, if $c_s>c$ at some density then we set $c_s=c$ at that density.  Unlike the piecewise polytropic parameterization, the spectral parameterization with these ranges of $\gamma_i$ cannot easily simulate a phase transition.  However, in other ways it is more flexible, with fewer parameters, than the piecewise polytropic model.

\subsubsection{Models using Gaussian processes}
\label{sec:Gaussian}

A newer approach to EOS modeling using GPs was introduced in \citet{2019PhRvD..99h4049L} (see also the subsequent work in \citealt{2020PhRvD.101f3007E}, \citealt{2020PhRvD.101l3007L}, and \citealt{2020PhRvC.102e5803E}).  Because this method is currently less commonly used in the community than the parametric approach, we will summarize it in more detail.

The essence of GP estimation of a function $f$ (see \citealt{2006gpml.book.....R} for an excellent introduction) is that we would like to have the outcome of our analysis be the joint probability density for the function values $f({\vec x})$ at inputs ${\vec x}$ (which we can represent as $x_i$), which we assume to be a multivariate Gaussian distribution with means ${\vec\mu}=\mu_i$ and covariance matrix $\Sigma=\Sigma_{ij}$.  That is, we would ultimately like to obtain
\begin{equation}
f({\vec x})\sim {\cal N}({\vec\mu},\Sigma)\; ,
\end{equation}
where ${\cal N}$ indicates a normal distribution.

A key assumption made in GP estimation is that correlations between the function values $f_i$, $f_i$ at inputs $x_i$, $x_j$ can be represented using a covariance kernel $K$ that is a function of $x_i$ and $x_j$.  That is, if our joint distribution is
\begin{equation}
f_i|x_i\sim {\cal N}(\langle f_i\rangle,\Sigma(f_i,f_j))\; ,
\end{equation} 
where $\langle f_i\rangle$ is the expectation value for $f$ at $x_i$ and $\Sigma(f_i,f_j)$ is the covariance matrix, then we assume we can write $\Sigma(f_i,f_j)=K(x_i,x_j)$.  

It is reasonable to assume that when $x_i$ and $x_j$ are closer to each other, $K$ is larger.  Given that covariance matrices are symmetric, it also must be that $K(x_i,x_j)=K(x_j,x_i)$ for any $x_i$ and $x_j$.  A common and much more drastic simplification is to assume that $K$ depends on only the distance between the points: $K(x_i,x_j)=K(|x_i-x_j|)$.  We use a Gaussian kernel, which in this context is called a ``squared-exponential kernel":
\begin{equation}
K_{\rm se}(x,x^\prime)=\sigma^2\exp\left(-{(x-x^\prime)^2\over{2\ell^2}}\right)\; ,
\label{eq:kernel}
\end{equation}
which has the properties listed above.  Here the hyperparameter $\sigma$ determines the strength of the overall correlation and the hyperparameter $\ell$ gives the correlation length scale.  For example, $\ell\rightarrow 0$ would mean that all of the points are independent of each other, and $\sigma\rightarrow 0$ would mean that there is negligible variance.

\textit{Transformation of dependent variable.}---There is one additional important qualitative aspect of the analysis of EOS using GPs.  A true Gaussian has a domain of $-\infty$ to $+\infty$.  However, if we think about an equation of state as $\epsilon(P)$, then because $\epsilon$ is nonnegative it cannot range from $-\infty$ to $+\infty$.  We could instead use $\log(\epsilon)$ as a function of $\log(P)$, because if we express $\epsilon$ in some units, e.g., cgs units, $\log(\epsilon)$ can range from $-\infty$ to $+\infty$.  However, this approach is also sub-optimal because it could lead to unstable ($dp/d\epsilon<0$) or acausal ($dp/d\epsilon>c^2$) EOS.  In principle one could simply discard such EOS when they are drawn from the GP, but this would be inefficient.

Therefore, we follow \citet{2010PhRvD..82j3011L} and \citet{2019PhRvD..99h4049L} in defining a new variable:
\begin{equation}
\phi\equiv\ln\left(c^2{d\epsilon\over{dP}}-1\right)\; .
\end{equation}
Because the adiabatic speed of sound $c_s$ is given by $c_s^2=dP/d\epsilon$, at the boundary of thermodynamic stability ($c_s^2\rightarrow 0$), $\phi\rightarrow+\infty$, and at the boundary of causality ($c_s^2\rightarrow c^2$), $\phi\rightarrow-\infty$.  Thus if we construct a GP in $\phi(\log P)$, all values of $\phi$ correspond to stable and causal EOS.

The next choice we need to make is how closely we wish to follow tabulated EOS in our function $\phi(\log P)$.  \citet{2019PhRvD..99h4049L} define two approximate limits: the ``model-informed" EOS prior, in which the GP is strongly conditioned on a set of EOS proposed in the literature, and the ``model-agnostic" EOS prior, in which the GP is only loosely related to existing EOS proposals.  The model-agnostic approach is much less computationally demanding and is less biased by existing EOS, so it is the path that we follow.

We then need to choose the parameters that describe our GP.  \citet{2019PhRvD..99h4049L} note that it is advantageous to look for approximate trends in the function we wish to model (here $\phi(\log P)$) and then to produce a GP for the residuals.  We find from sets of tabulated EOS (e.g., those at the CompOSE website https://compose.obspm.fr/table/families/3/) that in the log pressure range of interest ($\log_{10} P ({\rm erg~cm}^{-3})\sim 32-36$) the trend is roughly linear.  We use $\phi_0=5.5-2.0(\log_{10}P-32.7)$ as our approximate trend, but construction of the GP does not require an exact fit.  For our kernel, we use the squared-exponential kernel of Equation~(\ref{eq:kernel}) and, based on the spread between the EOS above in the $\log_{10} P ({\rm erg~cm}^{-3})\sim 32-36$ range, we choose $\sigma=1$ and $\ell=1$.  

We note that the Gaussian process framework allows for substantial flexibility beyond our particular choice (see, e.g., \citealt{2020PhRvD.101f3007E}).  For example, although $\sigma=1$ and $\ell=1$ matches well most existing EOS in the density range that has a discernible impact on the maximum mass, radius, and tidal deformability of neutron stars, when combined with our $\phi_0=5.5-2.0(\log_{10}P-32.7)$ trend it leads to sound speeds approaching the speed of light at densities several times nuclear saturation density.  A different choice of hyperparameters, e.g., a larger $\sigma$, would not necessarily require such high sound speeds; see for example Figure~2 of \citet{2020PhRvD.101l3007L}.

\ 
\bigskip  

\subsection{EOS results}
\label{sec:results}

\begin{figure*}[ht!]
\vspace{-2.0truein}
\begin{center}
  \resizebox{1.0\textwidth}{!}{\includegraphics{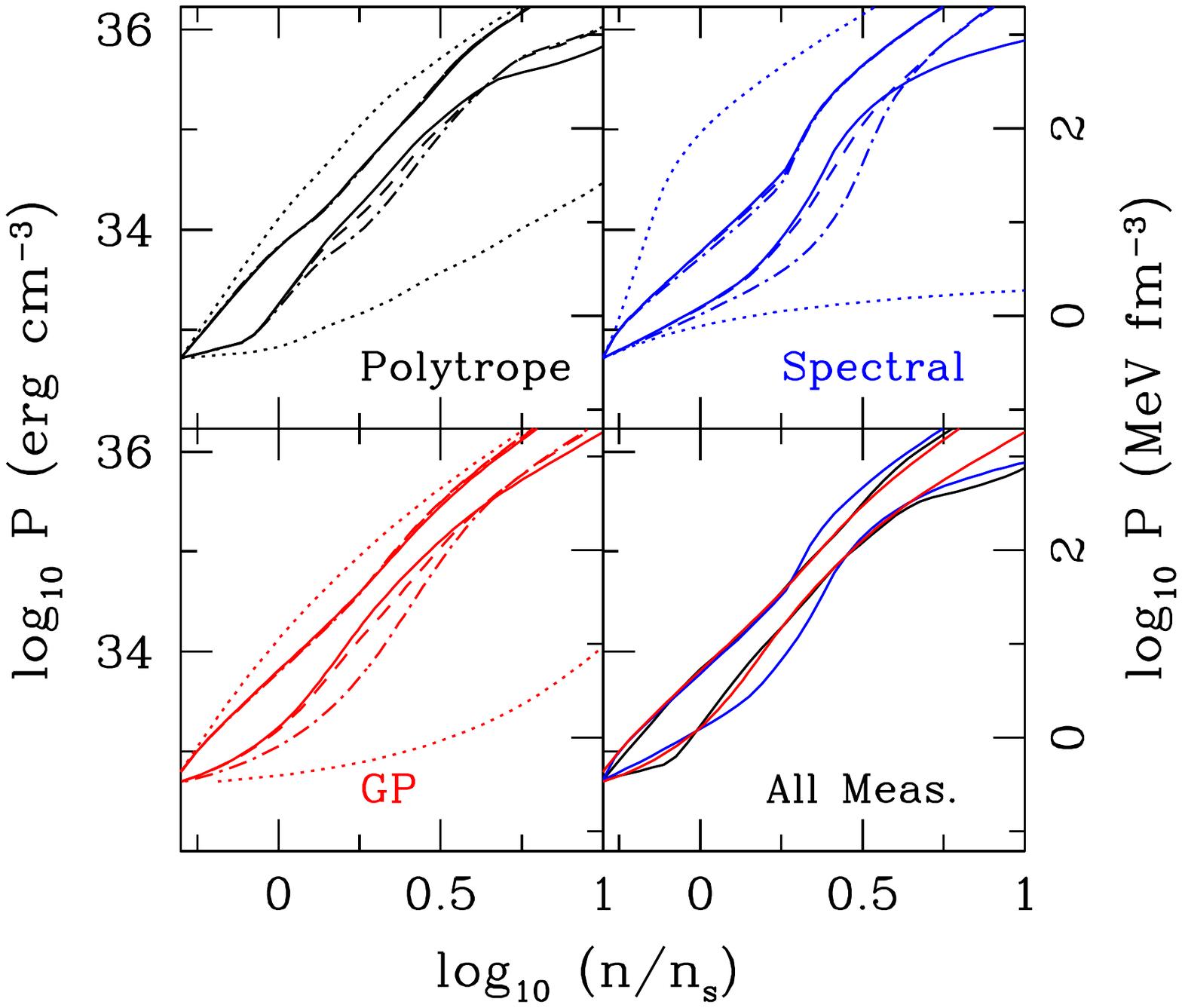}}
\vspace{-2.5truein}
   \caption{Comparison of the pressure ranges as a function of number density for the three EOS models described in the text, as a function of the data sets included in the analysis.  For each line type and color in each panel, the lower line shows the 5th percentile, and the upper line shows the 95th percentile, of the pressure as a function of density.  Colors indicate the EOS model used in the analysis: black for a piecewise polytrope (see the first part of Section~\ref{sec:parametric}), blue for the spectral model (see the second part of Section~\ref{sec:parametric}), and red for a model based on Gaussian processes (see Section~\ref{sec:Gaussian}).  We show the results for our priors (dotted lines in each panel); our priors plus symmetry energy measurements, the existence of three high-mass pulsars, and two tidal deformability upper limits (dash-dotted lines in each panel); those plus the mass-radius posteriors on PSR~J0030$+$0451 from \citet{2019ApJ...887L..24M} (dashed lines in each panel); and finally all of those plus the mass-radius measurement of PSR~J0740$+$6620 that we report here (solid lines in each panel).  The lower right panel shows the full set of constraints, including our PSR~J0740$+$6620 radius measurement, for each of our three EOS families.  Inclusion of the masses and radii of PSR~J0030$+$0451 and PSR~J0740$+$6620 tightens significantly the EOS models in the vicinity of $n/n_s \sim 1.5-3$ [$\log_{10} (n/n_s) \sim 0.2$--0.5].  This indicates that with the \textit{NICER} and \textit{XMM-Newton} data added, in this density range the posterior is now dominated by the data rather than by the priors.}
\label{fig:EOSall}
\end{center}
\end{figure*}

\begin{figure*}[ht!]
\vspace{-1.5truein}
\begin{center}
  \resizebox{1.0\textwidth}{!}{\includegraphics{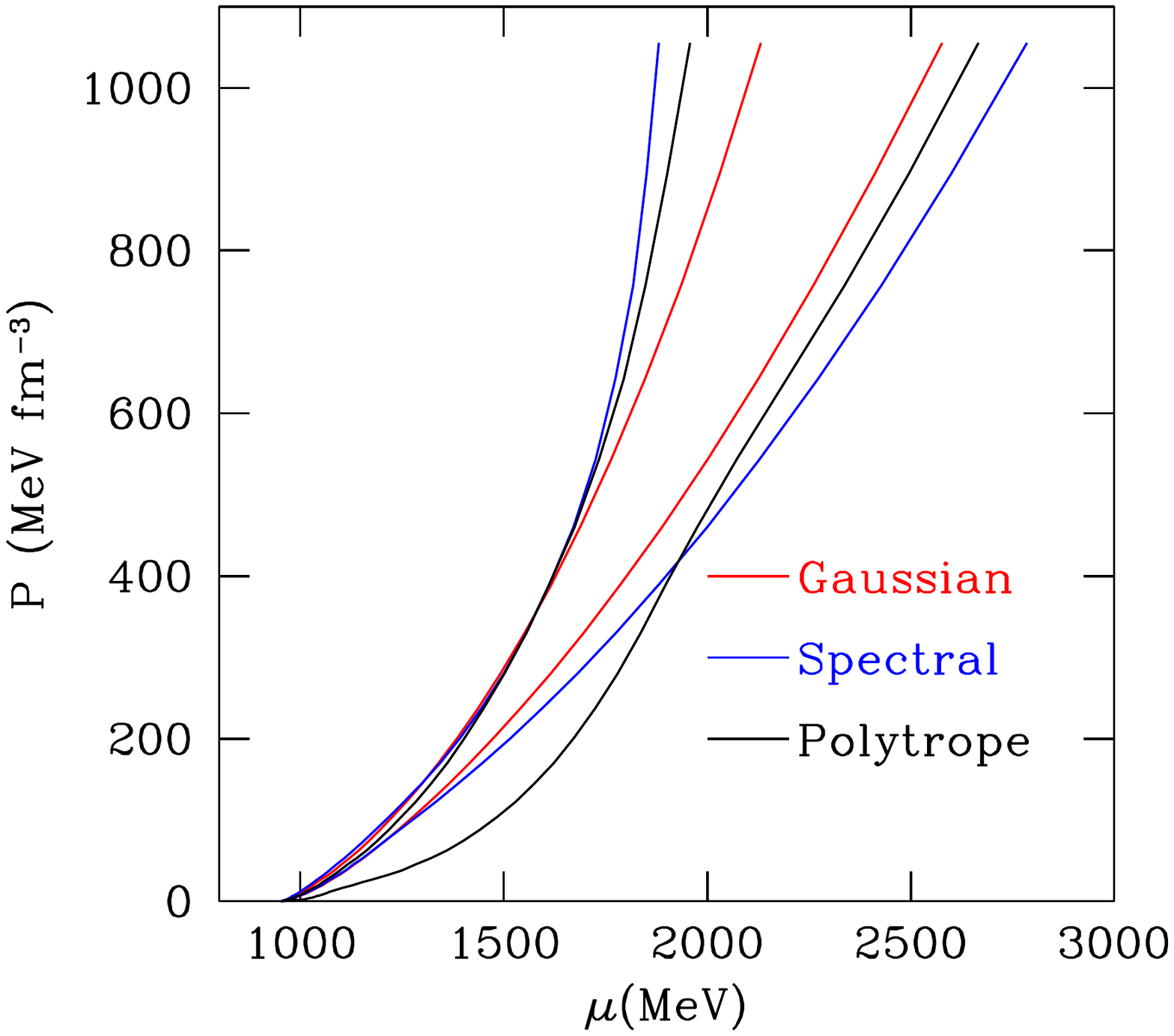}}
\vspace{-2.0truein}
   \caption{The 5th to 95th percentile range of the pressure as a function of the baryon chemical potential, including the neutron rest mass-energy, for all three of our models of the EOS: the red line is for the Gaussian process EOS, the blue line is for the spectral EOS, and the black line is for the piecewise polytropic EOS.  Here we include all measurements: symmetry energy, high-mass pulsars, tidal deformability from gravitational wave measurements, and measurements of the mass and radius of PSR~J0030$+$0451 and PSR~J0740$+$6620.}
\label{fig:mu}
\end{center}
\end{figure*}

\begin{figure*}[ht!]
\vspace{-0.1truein}
\begin{center}
  \resizebox{0.8\textwidth}{!}{\includegraphics{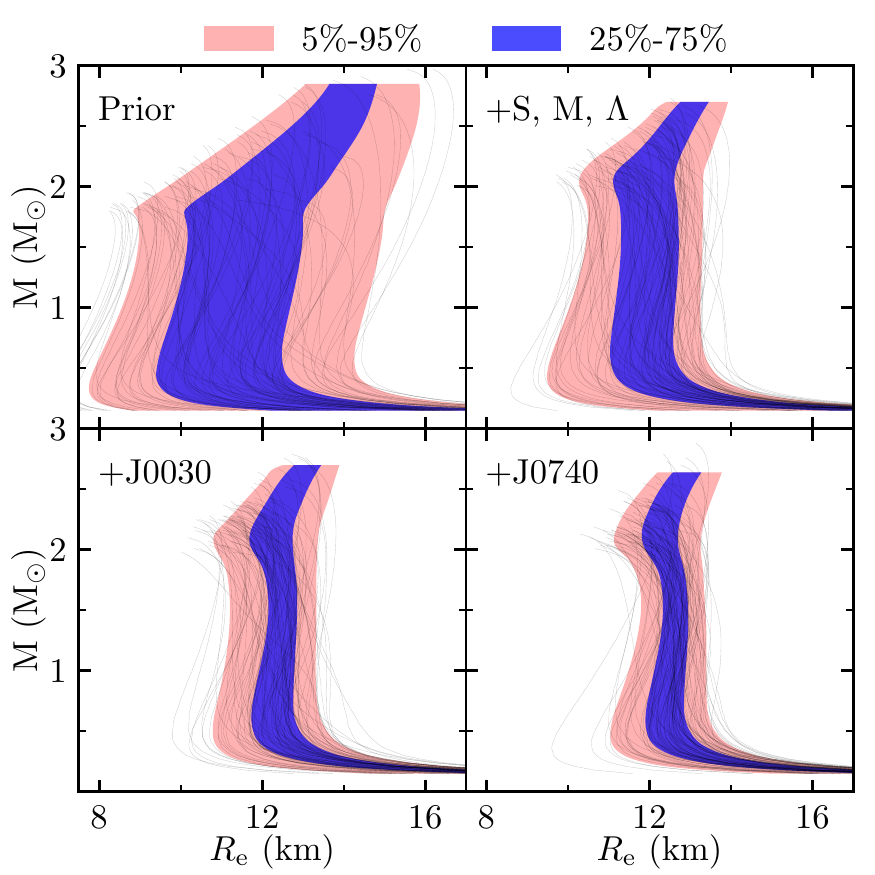}}
\vspace{+0.1truein}
   \caption{The 25th to 75th percentile range (blue shading) and the 5th to 95th percentile range (pink shading) of the radius as a function of mass, for nonrotating neutron stars.  The faint lines show the mass-radius relations of representative individual EOS (see text for details).  At each mass only EOS that can reach that mass contribute to the radius posterior; thus at high masses, which require hard EOS, the radii become larger.  We use the Gaussian process EOS model for the same progression of measurements as in Figure~\ref{fig:EOSall}, where $S$ is the symmetry energy, $M$ refers to the high masses of three pulsars, and $\Lambda$ indicates the gravitational wave measurements of tidal deformability for GW170817 and GW190425.  In each panel, we only include masses below the 95th percentile of the maximum mass for nonrotating neutron stars (see Figure~\ref{fig:Mmax} for a closer investigation of the maximum mass).  See Table~\ref{tab:mmaxrad} for details of how our radius bounds change between EOS models and when including different observations.  The agreement between the methods, particularly at the $\pm 1\sigma$ level, is another indication of improving convergence between models.}
\label{fig:RM}
\end{center}
\end{figure*}

\begin{figure*}[ht!]
\vspace{-1.5truein}
\begin{center}
  \resizebox{1.0\textwidth}{!}{\includegraphics{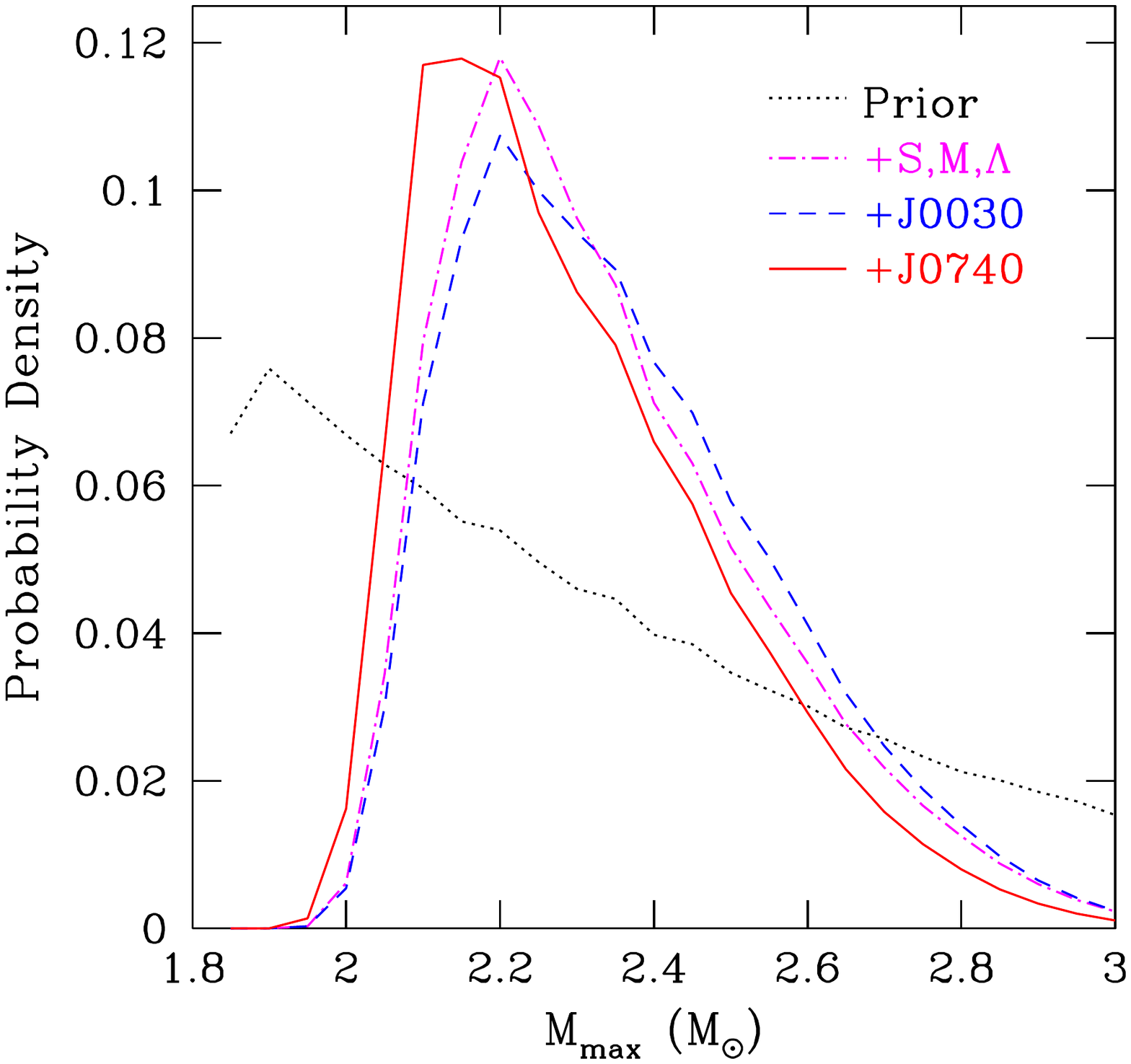}}
\vspace{-2.0truein}
   \caption{The posterior probability distribution for the maximum gravitational mass of a nonrotating neutron star, normalized in each case so that the integral of the probability density in the $M_{\rm max}=1.8-3.0~M_\odot$ range is 1.  The highest maximum mass allowed in our analysis was $3.0~M_\odot$.  We use the Gaussian process model for the same progression of measurements as in Figure~\ref{fig:EOSall}, where $S$ is the symmetry energy, $M$ refers to the high masses of three pulsars, and $\Lambda$ indicates the gravitational wave measurements of tidal deformability for GW170817 and GW190425.  See Table~\ref{tab:mmaxrad} for details of how our maximum mass bounds change between EOS models and when including different observations.}
\label{fig:Mmax}
\end{center}
\end{figure*}

\begin{deluxetable*}{c|c|c|c|c|c|c|c|c|c|c}
    \tablecaption{Summary of Maximum Mass and Radii at $1.4~M_\odot$ and $2.08~M_\odot$}
\tablewidth{0pt}
\tablehead{
      \colhead{EOS Model} & \colhead{Measurements} & \multicolumn{3}{c}{$M_{\rm max}~(M_\odot)$} & \multicolumn{3}{c}{$R_e(1.4~M_\odot)$, km }& \multicolumn{3}{c}{$R_e(2.08~M_\odot)$, km}   
    \label{tab:mmaxrad}
 }
\startdata
      \hline
                &               & $-1\sigma$ & Median & $+1\sigma$ & $-1\sigma$ & Median & $+1\sigma$ & $-1\sigma$ & Median & $+1\sigma$ \\
      \hline
      Gaussian  & $S,M,\Lambda$ & 2.12 & 2.27 & 2.52 & 10.92 & 12.23 & 12.93 & 10.80 & 11.97 & 12.86 \\
                & $+$J0030 & 2.13 & 2.30 & 2.54 & 11.88 & 12.51 & 13.02 & 11.39 & 12.27 & 12.95\\
                & $+$J0740 & 2.08 & 2.23 & 2.47 & 12.17 & 12.63 & 13.11 & 11.60 & 12.28 & 12.88\\
      \hline
      Spectral & $S,M,\Lambda$ & 2.28 & 2.55 & 2.88 & 10.41 & 11.43 & 12.40 & 10.79 & 11.81 & 12.97\\
                & $+$J0030 & 2.39 & 2.78 & 2.93 & 11.52 & 12.22 & 12.67 & 11.56 & 12.77 & 13.11\\
                & $+$J0740 & 2.23 & 2.74 & 2.92 & 11.79 & 12.30 & 12.84 & 11.83 & 12.78 & 13.11\\
      \hline 
      Piecewise & $S,M,\Lambda$ & 2.13 & 2.37 & 2.64 & 11.26 & 12.32 & 12.89 & 11.04 & 12.21 & 12.90\\
      Polytrope & $+$J0030 & 2.14 & 2.41 & 2.65 & 11.95 & 12.47 & 12.94 & 11.42 & 12.35 & 12.94\\
                & $+$J0740 & 2.09 & 2.27 & 2.61 & 12.16 & 12.56 & 13.01 & 11.67 & 12.36 & 12.91\\
      \hline 
\enddata
\tablecomments{Maximum gravitational masses and equatorial circumferential radii at $M=1.4~M_\odot$ and $M=2.08~M_\odot$ (the best estimate of the mass of PSR~J0740$+$6620), all at $\pm 1\sigma$ for nonrotating stars, inferred using our three EOS frameworks with three different sets of measurements.  The $S,M,\Lambda$ set includes constraints on the symmetry energy, the high masses of three pulsars, and the two LIGO/Virgo tidal deformability measurements.  The $+$J0030 set also includes the \citet{2019ApJ...887L..24M} measurement of the radius and mass of PSR~J0030$+$0451.  The $+$J0740 data adds our measurement of the radius of PSR~J0740$+$6620.  We see that the radius estimates tighten with addition of more data, and in particular that the final $\pm 1\sigma$ radius range for a $2.08~M_\odot$ star, spanning all three EOS frameworks ($11.60-13.11$~km), is very similar to the final $\pm 1\sigma$ radius range for a $1.4~M_\odot$ star, spanning all three EOS frameworks ($11.79-13.11$~km).}
\end{deluxetable*}

The constraints we obtain on the EOS at different densities depend both on the EOS model and priors and on the data sets that we include.  Here we present the EOS constraints for each of our three models, with progressive incorporation of more data.  We also plot the distributions of mass versus radius and the maximum mass for nonrotating neutron stars for different amounts of data and using the Gaussian process model.  For a given EOS and central density, we solve the Tolman-Oppenheimer-Volkoff equation \citep{1939PhRv...55..374O,1939PhRv...55..364T} to obtain the mass and radius.  Thus for the purpose of calculating stellar structure we implicitly treat the star as nonrotating.  This is an excellent approximation at the 346.53~Hz rotation frequency of this pulsar \citep{2020NatAs...4...72C}, because this is only $\sim 20\%$ of the mass-shedding frequency and thus the expected deviation from the structure of a nonrotating star is much smaller than our measurement precision (e.g., see Figure~13 of \citealt{2019ApJ...887L..24M}; at this mass and rotation frequency, the radius increase is at most 0.2~km compared with a nonrotating star).

In Figure~\ref{fig:EOSall} we show, for each EOS model, the middle 90\% range for the pressure as a function of number density when we incorporate successively more constraining measurements.  The upper left hand panel shows the results using the piecewise polytropic model, the upper right hand panel shows the results using the spectral model, and the lower left hand panel shows the results using the GP model.  Our first constraints (dotted lines) use only the prior for each EOS model.  The next constraints include a Gaussian prior of $S=32\pm 2$~MeV for the symmetry energy at nuclear saturation density (see \citealt{2012PhRvC..86a5803T}; note that the symmetry energy is the difference between the total energy per nucleon of pure neutron matter, which is close to $\epsilon/n$ because our matter is close to pure neutrons, and the energy per nucleon of symmetric nuclear matter), the existence of three high-mass pulsars \citep{2018ApJ...859...47A,2013Sci...340..448A,2020NatAs...4...72C}, and the tidal deformability posteriors from gravitational wave observations of GW170817 \citep{2017PhRvL.119p1101A,2018PhRvL.121p1101A,2018PhRvL.121i1102D} and GW190425 \citep{2020ApJ...892L...3A}.  Other constraints on the symmetry energy are possible; for example, \citet{2020PhRvL.125t2702D} find a tighter range for the symmetry energy of $S=31.7\pm 1.1$~MeV and analyses of PREX-II data \citep{2021arXiv210210767A,2021arXiv210103193R,2021arXiv210205267Y} suggest a significantly higher symmetry energy of $S\approx 38\pm 4$~MeV.  The third set of constraints also include the mass-radius posteriors from the \citet{2019ApJ...887L..24M} analysis of \textit{NICER} data on PSR~J0030$+$0451.  Finally, the last set of constraints also includes our mass-radius posteriors for PSR~J0740$+$6620.  The bottom right panel collects the final set of constraints for all three EOS models.  In each case, we only display the results for our fit to both the \textit{NICER} and the \textit{XMM-Newton} data using the nominal \textit{XMM-Newton} calibration.  When we include our mass-radius posteriors for PSR~J0740$+$6620, we use only the masses of PSR~1614--2230 and PSR~J0348$+$0432 in our high-pulsar-mass data, i.e., we do not double-count the high mass of PSR~J0740$+$6620.

Several trends are evident in this figure:

\begin{enumerate}

\item The 5\%-95\% pressure range at densities $n\approx \left(1.5-3\right)n_s$, after incorporation of all measurements, depends relatively weakly on the EOS model.  This is encouraging, because it means that for this density range the posterior is now dominated by the data rather than by the priors.

\item This convergence of constraints requires the \textit{NICER} data on PSR~J0030$+$0451 and the \textit{NICER} and \textit{XMM-Newton} data on PSR~J0740$+$6620.  Without these data, the pressure ranges differ significantly among the three models.

\item As expected, there is much less convergence at higher densities ($\gtorder 5n_s$) and at lower densities ($\ltorder 1.5n_s$).  This is because low and high densities contribute less to the radius, mass, or tidal deformability of neutron stars.  See \citet{2007PhR...442..109L,2016PhR...621..127L,2020arXiv200906441D,2020ApJ...899....4X} for the detailed theoretical context.

\item Finally, we note that the constraining power of our measurement of PSR~J0740$+$6620 comes entirely from the strong lower limit on its radius.  Other measurements, particularly the nondetection of tidal deformability from GW170817, already strongly exclude radii as large as even our $+1\sigma$ value of $>16$~km.  However, the evidence that the radius is $\gtorder 12$~km at $1\sigma$, and $\gtorder 11$~km at $2\sigma$, adds significantly to our understanding of dense matter. 

\end{enumerate}

In Figure~\ref{fig:mu} we present another view of the equation of state, after all measurements are incorporated.  Here we show the pressure versus the baryon chemical potential $\mu=(\epsilon+P)/n$ (including the baryon rest mass-energy) for all three of our EOS models.

In Figure~\ref{fig:RM} we present the 5th to 95th percentile range, and the 25th to 75th percentile range, of the radius as a function of mass, using our Gaussian process model, for the same set of measurements as in Figure~\ref{fig:EOSall}.  In this figure we also plot (faint lines) representative mass-radius relations drawn from our set of EOS, where in each panel the probability of drawing an individual EOS is proportional to its statistical weight given the measurements considered in that panel.  Here we only plot masses below the 95th percentile of the maximum gravitational mass of a nonrotating neutron star, given the measurements being considered.  The radius and mass of PSR~J0030$+$0451, and our current measurement of the radius of PSR~J0740$+$6620, tighten the allowed radius significantly at all neutron star masses greater than $1.0~M_\odot$.  See Table~\ref{tab:mmaxrad} for details.  In particular, the $\pm 1\sigma$ radius range at $M=1.4~M_\odot$, spanning all three EOS frameworks, is just $11.79-13.11$~km, or $\pm 5.3$\%.  In addition, although the direct measurement that we report of the radius of PSR~J0740$+$6620 has a $\pm 1\sigma$ range of $\approx 12.2-16.3$~km, when we include other measurements the $\pm 1\sigma$ range, spanning all three EOS frameworks, narrows to $11.60-13.11$~km at $M=2.08~M_\odot$.  This is dramatically improved compared with the prior $\pm 1\sigma$ range of $10.47-14.27$~km over all of our EOS frameworks.

In Figure~\ref{fig:Mmax} we see that the maximum mass range shifts slightly when we include our radius measurement of PSR~J0740$+$6620; see Table~\ref{tab:mmaxrad} for details.  The maximum mass range still depends on the EOS model.  However, consider for context the lighter object in the gravitational wave coalescence GW190814, which has a best mass estimate of $2.59~M_\odot$ \citep{2020ApJ...896L..44A}.  Using our Gaussian process EOS model, without our PSR~J0740$+$6620 measurement, there is a 12.1\% probability that this object is a slowly rotating neutron star.  With our PSR~J0740$+$6620 measurement, the probability drops to 7.5\%.

\section{CONCLUSIONS}
\label{sec:conclusions}

\textit{NICER} measurements of PSR~J0030$+$0451 and PSR~J0740$+$6620, complemented with \textit{XMM-Newton} measurements of PSR~J0740$+$6620, have reduced the uncertainty about the EOS of high-density, cold catalyzed matter.  This is especially true in the density range $n\sim 1.5-3n_s$.  These two radius measurements also shift slightly the credible region for the maximum gravitational mass of a nonrotating neutron star.  

Most importantly, our new measurement narrows significantly the inferred radius range for $1.4~M_\odot$ neutron stars.  For instance, prior to our measurement, when we used our Gaussian process EOS model, inclusion of nuclear data, information about neutron star tidal deformability from the gravitational wave event GW170817, and our previous mass and radius measurement of PSR~J0030$+$0451 had constrained the radius to $11.2-13.3$~km at 90\% credibility and $11.9-13.0$~km at 68\% credibility.  When we also include our PSR~J0740$+$6620 measurement, these ranges shrink to $11.8-13.4$~km (90\%) and $12.2-13.1$~km (68\%).  That is, as expected, the main effect of our PSR~J0740$+$6620 measurement is to increase the lower limit to the radius.  Indeed, even when we consider the radius range spanned by all three of our EOS models, the range is $11.8-13.1$~km at 68\% credibility, for a fractional uncertainty of $\pm 5.3$\%.  This is broadly consistent with previous radius estimates based both on nuclear theory and tidal deformabilities inferred from gravitational wave observations \citep{2017PhRvL.119p1101A,2018PhRvL.121i1102D,2020ApJ...892L...3A} and on the \textit{NICER} mass and radius estimates for PSR~J0030$+$0451 \citep{2019ApJ...887L..24M,2019ApJ...887L..21R}: the total radius range spanned by all the computed $\pm 1\sigma$ radius bounds, prior to our measurement of the radius of PSR~J0740$+$6620, for $1.4~M_\odot$ neutron stars is $\approx 11-13.5$~km (e.g., \citealt{2020EPJST.229.3615A,2020Sci...370.1450D,2020PhRvC.102e5803E,2020ApJ...901..155G,2020ApJ...892...55J,2020arXiv200912571L,2020PhRvD.102b3021Z,2021arXiv210305408H,2021arXiv210315119L,2021Symm...13..144S,2021arXiv210406141S}).

Data are still being taken actively; for example, it is likely that by the end of the \textit{NICER} mission the exposure times on PSR~J0030$+$0451 and PSR~J0740$+$6620 will roughly double compared with the times used in current analyses, and an analysis of \textit{NICER} data on PSR~J0437$-$4715 (which has the highest X-ray flux of any non-accreting neutron star) is in progress.  This gives reason for optimism that future \textit{NICER} observations, and in the coming years more gravitational-wave observations of tidal deformability, will provide even clearer insight into a realm of matter which cannot be explored in laboratories.

\acknowledgments

This work was supported by NASA through the \textit{NICER} mission and the Astrophysics Explorers Program.
The authors acknowledge the University of Maryland supercomputing resources\break
(http://hpcc.umd.edu) that were made available for conducting the research reported in this paper.  The authors also thank Felix Fuerst for facilitating the \textit{XMM-Newton} director's discretionary time observations of PSR~J0740+6620, and Brad Cenko for early \textit{Swift} observations.  MCM thanks the Radboud Excellence Initiative for supporting his stay at Radboud University, and thanks Gordon Baym, Cecilia Chirenti, and Sanjay Reddy for comments on an earlier version of this manuscript.  WCGH appreciates use of computer facilities at the Kavli Institute for Particle Astrophysics and Cosmology and acknowledges support through grant 80NSSC20K0278 from NASA.  TTP is a NANOGrav Physics Frontiers Center Postdoctoral Fellow funded by the National Science Foundation award number 1430284.  IHS is supported by an NSERC Discovery Grant and by the Canadian Institute for Advanced Research.  Support for HTC was provided by NASA through the NASA Hubble Fellowship Program grant \#HST-HF2-51453.001 awarded by the Space Telescope Science Institute, which is operated by the Association of Universities for Research in Astronomy, Inc., for NASA, under contract NAS5-26555.  SMM thanks the Natural Sciences and Engineering Research Council of Canada for funding.  Portions of this work performed at NRL were funded by NASA.  The authors acknowledge the use of NASA's Astrophysics Data System (ADS) Bibliographic Services and the arXiv.

\facility{\textit{NICER} (\citealt{2016SPIE.9905E..1HG}), \textit{XMM-Newton} (\citealt{2001A&A...365L..18S,2001A&A...365L..27T}), \textit{Swift} (\citealt{2004ApJ...611.1005G})}

\software{emcee (\citealt{2013PASP..125..306F}), MultiNest (\citealt{2009MNRAS.398.1601F}), Python and NumPy (\citealt{2007CSE.....9c..10O}), Matplotlib (\citealt{2007CSE.....9...90H}), Cython (\citealt{2011CSE....13b..31B}), schwimmbad (\citealt{schwimmbad}), and the \textit{XMM-Newton} Scientific Analysis System (SAS, \citealt{2004ASPC..314..759G})}

\newpage
\appendix

\section{Posterior Distributions from Analysis of Synthetic J0740-like \textit{NICER} and \textit{XMM-Newton} Data}
\label{sec:corner-plot-Synthetic}

\begin{deluxetable*}{crrrrrr}[h]
    \tablecaption{Fits to synthetic \textit{NICER} and \textit{XMM-Newton} data}
\tablewidth{0pt}
\tablehead{
      \colhead{Parameter} & \colhead{Median} & \colhead{$-1\sigma$} & \colhead{$+1\sigma$} & \colhead{$-2\sigma$}  & \colhead{$+2\sigma$} & \colhead{Assumed value}
   \label{tab:crediblesynthetic}
}
\startdata
      \hline
$R_{e}$ (km)& 13.933 & 12.344 & 17.396 & 11.342 & 22.215 & 13.378\\ 
\hline 
$GM/c^2R_e$&   0.221  & 0.177  & 0.248  & 0.139  & 0.267  & 0.230\\ 
\hline 
$M~(M_\odot$)&   2.081  & 1.991  & 2.174  & 1.895  & 2.261 & 2.080\\ 
\hline 
$\theta_{\rm c1}$ (rad)&  1.618  & 0.892  & 2.341  & 0.583  & 2.602 & 1.820\\ 
\hline 
$\Delta \theta_1$ (rad)&   0.121  & 0.078  & 0.180  & 0.051  & 0.259 & 0.061\\ 
\hline 
$kT_{\rm eff,1}$ (keV)&   0.085  & 0.076  & 0.097  & 0.068  & 0.111 & 0.107\\ 
\hline 
$\theta_{\rm c2}$ (rad)&  1.574  & 0.881  & 2.313  & 0.586  & 2.604 & 2.280\\ 
\hline 
$\Delta \theta_2$ (rad)&  0.121  & 0.078  & 0.179  & 0.050  & 0.262 & 0.086\\ 
\hline 
$kT_{\rm eff,2}$ (keV)&  0.085  & 0.076  & 0.097  & 0.068  & 0.111 & 0.100\\ 
\hline 
$\Delta \phi_2$ (cycles)&  0.554  & 0.422  & 0.578  & 0.409  & 0.592 & 0.422\\ 
\hline 
$\theta_{\rm obs}$ (rad)&   1.526  & 1.467  & 1.585  & 1.444  & 1.610 & 1.463\\ 
\hline 
$N_H~(10^{20}~\rm cm^{-2}$)&  3.067  & 1.377  & 4.389  & 0.302  & 4.915 & 0.226\\ 
\hline 
$d$ (kpc)&  1.211  & 1.025  & 1.400  & 0.840  & 1.598 & 1.099\\ 
\hline 
\enddata
\tablecomments{The one-dimensional credible intervals and best fit (median) values for each of the parameters in a pulse waveform model with two possibly different uniform circular spots, obtained by jointly fitting the model to synthetic \textit{NICER} data (in channels 30--123) and synthetic \textit{XMM-Newton} data that together mimic the actual \textit{NICER} and \textit{XMM-Newton} data. The rightmost column lists the parameter values that were used to generate the synthetic data. These values were taken from a model that gives a good fit to the actual PSR~J0740$+$6620 data. The synthetic data were then generated by Poisson draws from the fluxes predicted by that model, including the background fluxes it predicts. Consistent with statistical expectations, 6 of the 12 $\pm 1\sigma$ credible intervals, 11 of the 12 $\pm 2\sigma$ credible regions, and all (12 of 12) of the $\pm 3\sigma$ credible regions contain the values of the model parameters that were assumed in generating the synthetic data.}
\end{deluxetable*}

\newpage

\begin{figure*}[ht!]
\begin{center}
\vspace*{-0.1truein}
  \resizebox{0.9\textwidth}{!}{\includegraphics{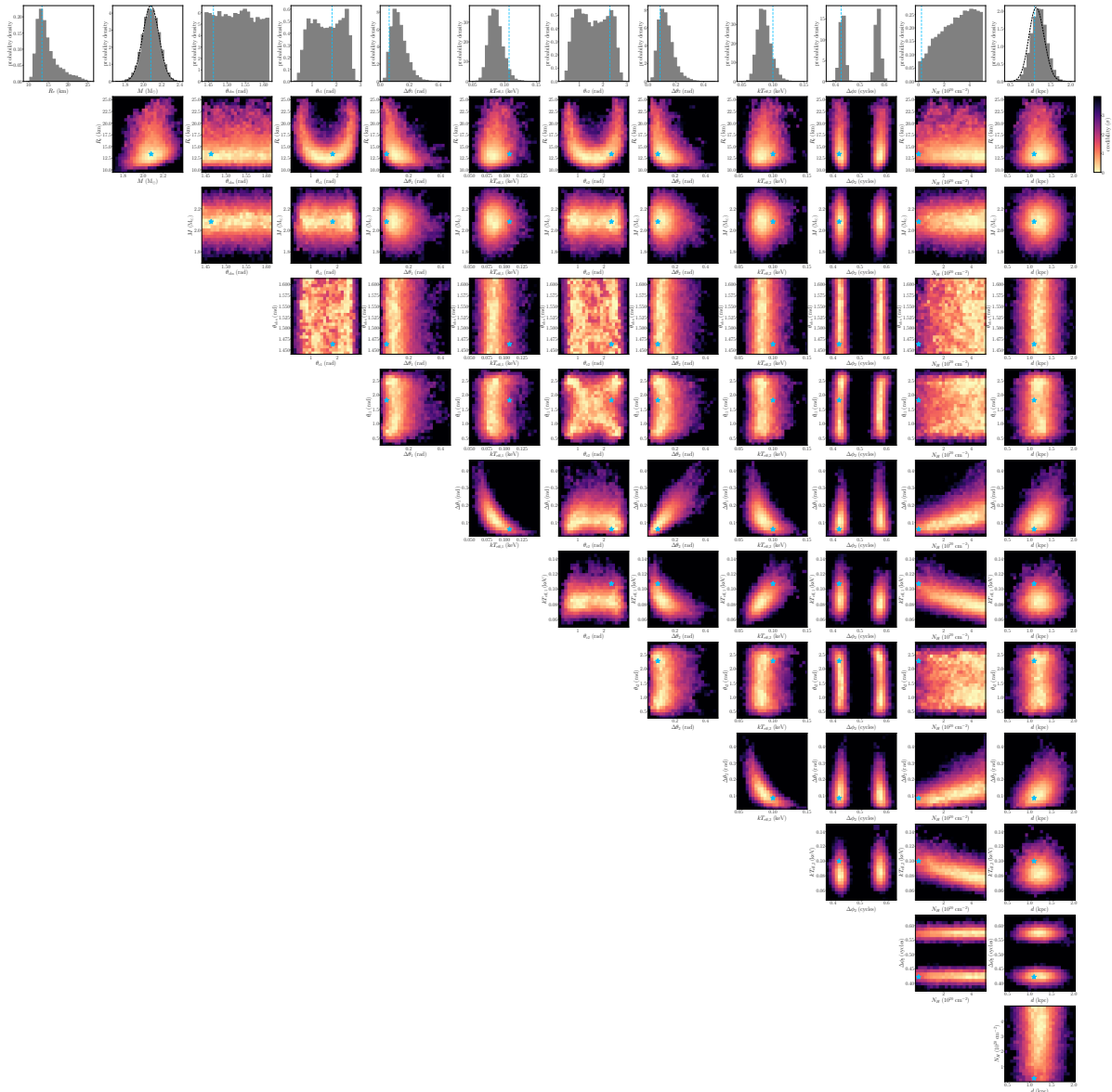}}
\vspace{-0.2truein}
   \caption{Posterior probability density distributions for a model with two, uniform, circular spots that is fit jointly to the synthetic \textit{NICER} and \textit{XMM-Newton} data.  The dotted lines in the one-dimensional plots for gravitational mass and distance indicate the priors that we applied.  The vertical dashed lines in the one-dimensional plots indicate the value of the associated parameter that was assumed in the construction of the synthetic data.  Similarly, the star symbols in the two-dimensional plots show the assumed values of both parameters.  In this and the other corner plots in the appendices, brighter colors in the two-dimensional plots indicate higher posterior probability densities.}
\label{fig:full-posteriors-for-synthetic}
\end{center}
\end{figure*}
\newpage

\section{Posterior Distributions from Analysis of Only \textit{NICER} Data}
\label{sec:corner-plot-NICERonly}

\begin{deluxetable*}{crrrrrr}[h]
    \tablecaption{Fits to only \textit{NICER} data}
\tablewidth{0pt}
\tablehead{
      \colhead{Parameter} & \colhead{Median} & \colhead{$-1\sigma$} & \colhead{$+1\sigma$} & \colhead{$-2\sigma$}  & \colhead{$+2\sigma$} & \colhead{Best fit}
   \label{tab:credibleNICERonly}
}
\startdata
      \hline
$R_{e}$ (km)& 11.512 & 10.382 & 13.380  & 9.642 & 16.256 & 11.008\\ 
\hline 
$GM/c^2R_e$&   0.266  & 0.229  & 0.293  & 0.188  & 0.308 & 0.278\\ 
\hline 
$M~(M_\odot$)&   2.072  & 1.978  & 2.159  & 1.885  & 2.247 & 2.071\\ 
\hline 
$\theta_{\rm c1}$ (rad)&  1.655  & 0.667  & 2.541  & 0.292  & 2.858 & 0.494\\ 
\hline 
$\Delta \theta_1$ (rad)&    0.329  & 0.205  & 0.541  & 0.132  & 0.899 & 0.328\\ 
\hline 
$kT_{\rm eff,1}$ (keV)&   0.072  & 0.064  & 0.079  & 0.057  & 0.087 & 0.078\\ 
\hline 
$\theta_{\rm c2}$ (rad)&   1.582  & 0.641  & 2.510  & 0.310  & 2.847 & 1.803\\ 
\hline 
$\Delta \theta_2$ (rad)&   0.326  & 0.204  & 0.539  & 0.132  & 0.854 & 0.208\\ 
\hline 
$kT_{\rm eff,2}$ (keV)&   0.072  & 0.064  & 0.079  & 0.057  & 0.087 & 0.079\\ 
\hline 
$\Delta \phi_2$ (cycles)&   0.527  & 0.441  & 0.560  & 0.422  & 0.578 & 0.550\\ 
\hline 
$\theta_{\rm obs}$ (rad)&   1.528  & 1.468  & 1.586  & 1.444  & 1.610 & 1.535\\ 
\hline 
$N_H~(10^{20}~\rm cm^{-2}$)&   1.833  & 0.599  & 3.414  & 0.094  & 4.667 & 0.858\\ 
\hline 
$d$ (kpc)&   1.213  & 1.025  & 1.407  & 0.843  & 1.596 & 1.086\\ 
\hline 
\enddata
\tablecomments{One-dimensional credible regions, and best fit, obtained by fitting a model with two possibly different uniform circular spots to only the \textit{NICER} data, using energy channels 30--123.  These credible regions may be compared with the regions in Table~\ref{tab:credibleNICERXMM}, where both \textit{NICER} and \textit{XMM-Newton} data are analyzed.  The analysis here was performed using a free background in each \textit{NICER} energy channel; as discussed in Section~\ref{sec:estimating-J0740-mass+radius}, future incorporation of reliable background models for \textit{NICER} will likely increase the \textit{NICER}-only radius to values more comparable with what we infer when we also use \textit{XMM-Newton} data.}
\end{deluxetable*}

\newpage 

\begin{figure*}[ht!]
\begin{center}
\vspace*{-0.1truein}
  \resizebox{0.9\textwidth}{!}{\includegraphics{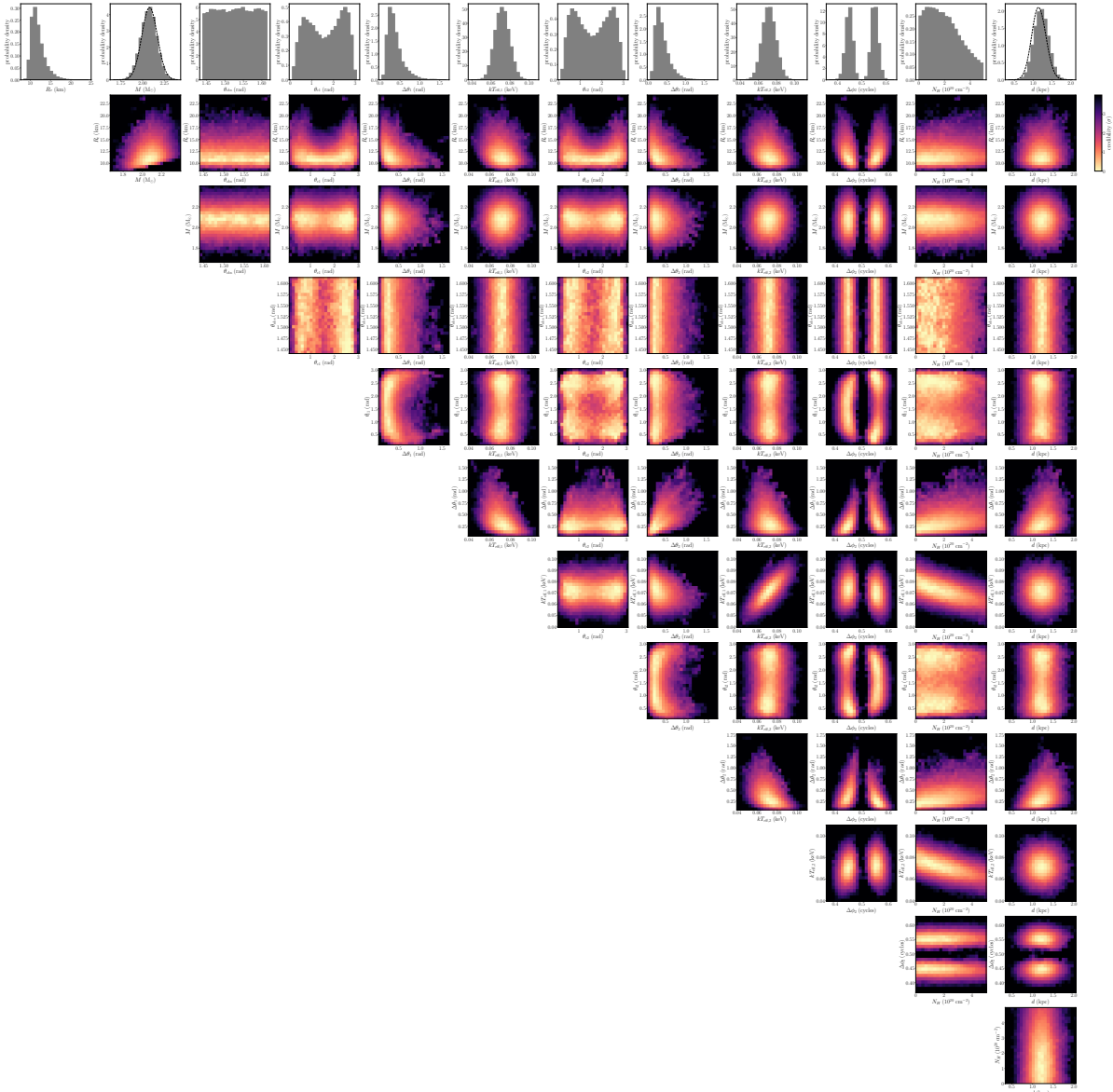}}
\vspace{-0.2truein}
   \caption{Posterior probability density distributions for a model with two, uniform, circular spots that is fit to only the \textit{NICER} data.  Note that because the spots do not overlap with each other, their roles can be swapped.  Moreover, because the observer inclination is close to the equator, there is a near-degeneracy between spots in the northern hemisphere and spots in the southern hemisphere.  The dotted lines in the one-dimensional plots for gravitational mass and distance indicate the priors that we applied.}
\label{fig:full-posteriors-for-NICER-only}
\end{center}
\end{figure*}
\newpage

\section{Posterior Distributions from Analysis of \textit{NICER} and \textit{XMM-Newton} Data}
\label{sec:corner-plot-NICERXMM}

\begin{deluxetable*}{crrrrrr}[h]
    \tablecaption{Fits to \textit{NICER} and \textit{XMM-Newton} data}
\tablewidth{0pt}
\tablehead{
      \colhead{Parameter} & \colhead{Median} & \colhead{$-1\sigma$} & \colhead{$+1\sigma$} & \colhead{$-2\sigma$}  & \colhead{$+2\sigma$} & \colhead{Best fit}
   \label{tab:credibleNICERXMM}
}
\startdata
      \hline
$R_{e}$ (km)& 13.713 & 12.209 & 16.326 & 11.243 & 20.051 & 13.823\\ 
\hline 
$GM/c^2R_e$& 0.222  & 0.187  & 0.249  & 0.151  & 0.267 & 0.221\\
\hline 
$M~(M_\odot$)& 2.062  & 1.971  & 2.152  & 1.883  & 2.242 & 2.067\\ 
\hline 
$\theta_{\rm c1}$ (rad)& 1.600  & 0.900  & 2.319  & 0.586  & 2.631 & 0.834\\ 
\hline 
$\Delta \theta_1$ (rad)& 0.098  & 0.065  & 0.145  & 0.044  & 0.219 & 0.087\\ 
\hline 
$kT_{\rm eff,1}$ (keV)& 0.094  & 0.083  & 0.105  & 0.073  & 0.118 & 0.098\\ 
\hline 
$\theta_{\rm c2}$ (rad)& 1.612  & 0.917  & 2.297  & 0.572  & 2.618 & 1.223\\ 
\hline 
$\Delta \theta_2$ (rad)& 0.096  & 0.064  & 0.143  & 0.042  & 0.213 & 0.066\\ 
\hline 
$kT_{\rm eff,2}$ (keV)& 0.094  & 0.083  & 0.106  & 0.073  & 0.120 & 0.103\\ 
\hline 
$\Delta \phi_2$ (cycles)& 0.558  & 0.422  & 0.579  & 0.409  & 0.592 & 0.577\\ 
\hline 
$\theta_{\rm obs}$ (rad)& 1.527  & 1.467  & 1.586  & 1.444  & 1.610 & 1.561\\ 
\hline 
$N_H~(10^{20}~\rm cm^{-2}$)& 1.137  & 0.315  & 2.549  & 0.043  & 4.174 & 0.006\\ 
\hline 
$d$ (kpc)& 1.215  & 1.027  & 1.404  & 0.852  & 1.593 & 1.151\\ 
\hline 
\enddata
\tablecomments{One-dimensional credible regions, and best fit, obtained by jointly fitting a model with two possibly different uniform circular spots to channels 30--123 of the \textit{NICER} data and to the \textit{XMM-Newton} data on PSR~J0740$+$6620.    These credible regions may be compared with the regions in Table~\ref{tab:credibleNICERonly}, where only \textit{NICER} data are analyzed.}
\end{deluxetable*}

\newpage

\begin{figure*}[ht!]
\begin{center}
\vspace*{-0.1truein}
  \resizebox{0.9\textwidth}{!}{\includegraphics{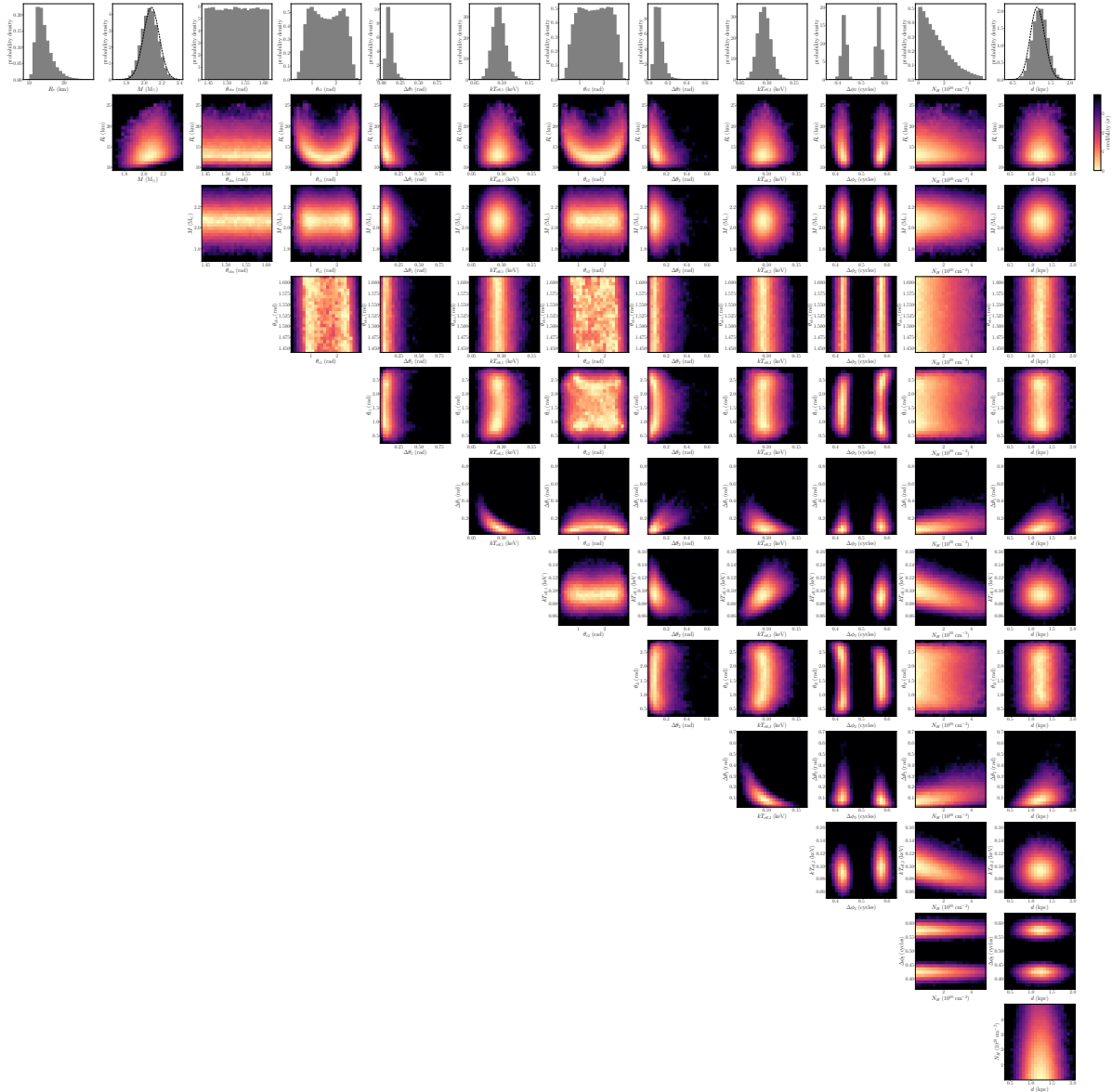}}
\vspace{-0.2truein}
   \caption{Posterior probability density distributions for a model with two, uniform, circular spots that is fit to the \textit{NICER} and \textit{XMM-Newton} data.  The dotted lines in the one-dimensional plots for gravitational mass and distance indicate the priors that we applied.}
\label{fig:full-posteriors-for-NICER-and-XMM}
\end{center}
\end{figure*}

\newpage

\section{Posterior Distributions from Analysis of \textit{NICER} and \textit{XMM-Newton} Data with an \textit{XMM-Newton} Calibration Parameter}
\label{sec:corner-plot-NICERXMMcalib}

\begin{deluxetable*}{crrrrrr}[h]
    \tablecaption{Fits to \textit{NICER} and \textit{XMM-Newton} data with a parameterized normalization for the \textit{XMM-Newton} calibration.}
\tablewidth{0pt}
\tablehead{
      \colhead{Parameter} & \colhead{Median} & \colhead{$-1\sigma$} & \colhead{$+1\sigma$} & \colhead{$-2\sigma$}  & \colhead{$+2\sigma$} & \colhead{Best fit}
   \label{tab:credibleNICERXMMnorm}
}
\startdata
      \hline
$R_{e}$ (km)& 13.705 & 12.153 & 16.298 & 11.140 & 20.199 & 13.524\\ 
\hline 
$GM/c^2R_e$&  0.223  & 0.187  & 0.250  & 0.151  & 0.270  & 0.227 \\
\hline 
$M~(M_\odot$)&  2.064  & 1.973  & 2.154  & 1.882  & 2.244 & 2.082\\ 
\hline 
$\theta_{\rm c1}$ (rad)&  1.635  & 0.895  & 2.324  & 0.575  & 2.634 & 2.399\\ 
\hline 
$\Delta \theta_1$ (rad)&  0.097  & 0.064  & 0.144  & 0.042  & 0.214 & 0.106\\ 
\hline 
$kT_{\rm eff,1}$ (keV)&  0.095  & 0.084  & 0.107  & 0.074  & 0.120 & 0.098\\ 
\hline 
$\theta_{\rm c2}$ (rad)&  1.647  & 0.910  & 2.327  & 0.573  & 2.632 & 1.876\\ 
\hline 
$\Delta \theta_2$ (rad)&  0.097  & 0.065  & 0.143  & 0.044  & 0.212 & 0.076\\ 
\hline 
$kT_{\rm eff,2}$ (keV)&  0.095  & 0.084  & 0.106  & 0.074  & 0.119 & 0.102\\ 
\hline 
$\Delta \phi_2$ (cycles)&  0.551  & 0.422  & 0.578  & 0.409  & 0.591 & 0.571\\ 
\hline 
$\theta_{\rm obs}$ (rad)&  1.525  & 1.467  & 1.586  & 1.444  & 1.610 & 1.536\\ 
\hline 
$N_H~(10^{20}~\rm cm^{-2}$)&  1.081  & 0.293  & 2.489  & 0.040  & 4.116 & 0.033\\ 
\hline 
$d$ (kpc)&  1.209  & 1.023  & 1.403  & 0.845  & 1.599 & 1.206\\ 
\hline
$A_{\rm XMM}$ &  0.969  & 0.919  & 1.043  & 0.902  & 1.090 & 0.917\\
\hline 
\enddata
\tablecomments{One-dimensional credible regions, and best fit, obtained by jointly fitting a model with two possibly different uniform circular spots to channels 30--123 of the \textit{NICER} data and to the \textit{XMM-Newton} data on PSR~J0740$+$6620.  Here we add a parameter, $A_{\rm XMM}$, which is the energy-independent ratio of the \textit{XMM-Newton} effective area to its nominal value.  These credible regions are nearly identical to those in Table~\ref{tab:credibleNICERXMM}, where we assumed the nominal effective area.}
\end{deluxetable*}

\newpage

\begin{figure*}[ht!]
\begin{center}
\vspace*{-0.1truein}
  \resizebox{0.9\textwidth}{!}{\includegraphics{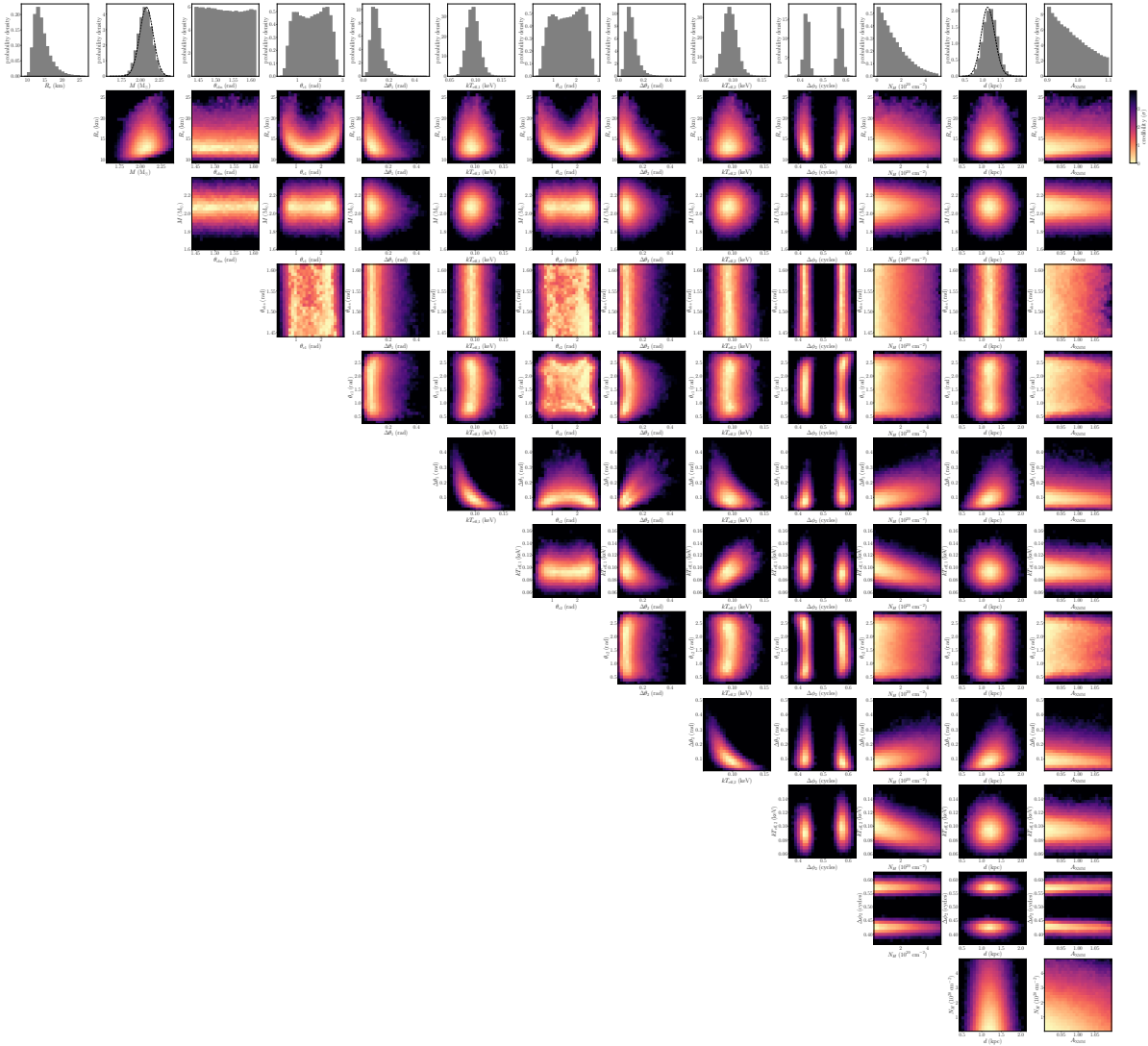}}
\vspace{-0.2truein}
   \caption{Posterior probability density distributions for a model with two, uniform, circular spots that is fit to the \textit{NICER} and \textit{XMM-Newton} data, where we allow the \textit{XMM-Newton} effective area to be multiplied by an energy-independent parameter $A_{\rm XMM}$.  The dotted lines in the one-dimensional plots for gravitational mass and distance indicate the priors that we applied.}
\label{fig:full-posteriors-for-NICER-and-XMMcalib}
\end{center}
\end{figure*}

\newpage
\bibliography{bibfileJ0740}

\end{document}